\documentclass[11pt]{article}
\pdfoutput=1 
\usepackage{microtype}
\usepackage{mathtools}
\usepackage{booktabs}
\usepackage[english]{babel}
\usepackage{amsmath,amssymb,amsbsy,amstext, amsthm, simplewick, amsfonts}
\usepackage{empheq}
\usepackage[most]{tcolorbox}
\usepackage{graphicx}
\usepackage[small]{caption}
\usepackage{upgreek}
\usepackage{framed}
\usepackage{wrapfig}
\usepackage{multirow}
\usepackage{bbm}
\usepackage[numbers,sort&compress]{natbib}
\usepackage[svgnames,dvipsnames,x11names]{xcolor}
\usepackage[utf8x]{inputenc}
\usepackage{selinput}
\usepackage{bm}
\usepackage{float}
\usepackage[top=2.5cm, bottom=2.5cm, left=3cm, right=3cm]{geometry}
\usepackage{yfonts}
\usepackage{caption}
\usepackage{subcaption}
\usepackage{sidecap}
\usepackage{longtable}
\usepackage{anyfontsize}
\usepackage{hyperref}
\usepackage{cleveref}

\usepackage{epstopdf}
\usepackage{cancel}
\usepackage{tcolorbox}
\usepackage{appendix}
\usepackage{pgfornament}

\DeclareFontFamily{OT1}{rsfs10}{}
\DeclareFontShape{OT1}{rsfs10}{m}{n}{ <-> rsfs10 }{}
\DeclareMathAlphabet{\mathscript}{OT1}{rsfs10}{m}{n}
\newcommand{\be}{\begin{equation}}
\newcommand{\ee}{\end{equation}}
\newcommand{\bea}{\begin{eqnarray}}
\newcommand{\eea}{\end{eqnarray}}
\newcommand{\bmat}{\begin{bmatrix}}
\newcommand{\emat}{\end{bmatrix}}
\newcommand{\beq}{\begin{equation}}
\newcommand{\eeq}{\end{equation}}
\newcommand{\beqa}{\begin{eqnarray}}
\newcommand{\eeqa}{\end{eqnarray}}
\newcommand{\beqar}{\begin{eqnarray*}}
\newcommand{\eeqar}{\end{eqnarray*}}

\newcommand{\bbibitem}[1]{\bibitem{#1}\marginpar{#1}}

\def\Label#1{\label{#1}%
  \smash{\hbox to0pt{\raise1ex\hbox{\tiny[#1]}\hss}}}
\def\noLabels{\let\Label=\label}
\def\nobbibitem{\let\bbibitem=\bibitem}

\newcommand{\hypergeom}[2]{
  \mathbin{_{#1}{\sf F}_{#2}} }
\makeatletter
\DeclareRobustCommand{\rcite}[1]{%
  \rcite@aux#1,\@nil{#1}%
}
\def\rcite@aux#1,#2\@nil#3{%
  \if\relax#2\relax
    Ref.~\cite{#3}%
  \else
    Refs.~\cite{#3}%
  \fi
}
\makeatother
\definecolor{greyish}{rgb}{.90,.90,.90}
\definecolor{greyish2}{rgb}{.96,.96,.96}
\usepackage{xcolor,colortbl}
\usepackage{tcolorbox}
\usepackage{imakeidx}
\numberwithin{equation}{section}
\makeindex
\begin{document}
\renewcommand{\thefootnote}{\fnsymbol{footnote}}
\vspace{0truecm}
\thispagestyle{empty}
\hfill
\begin{center}

{\fontsize{21}{18} \bf Clines and the Analytic Structure of \\[14pt]
Black Hole Perturbations}
\end{center}
\vspace{.15truecm}
 \vspace{.25cm}

\vspace{.15truecm}

\begin{center}

{\fontsize{13}{18}\selectfont
Maria J. Rodriguez $^{\rm a,b}$\footnote{\texttt{majo.rodriguez.b@gmail.com}} and 
Luis Fernando Temoche \ $^{\rm a}$\footnote{\texttt{{lfth93@gmail.com}}}}
\end{center}

\vspace{.4truecm}
\begin{scriptsize}

 \centerline{{\it ${}^{\rm a}$Department of Physics, Utah State University, Logan, UT 84322, USA}}
\vspace{.05cm}

\centerline{{\it ${}^{\rm b}$Instituto de F\'\i sica Te\'orica UAM-CSIC,
 Madrid, 28049, Spain}}

\end{scriptsize}

 \vspace{.25cm}

\vspace{.3cm}
\begin{abstract}
\noindent

We revisit black hole perturbations through Heun differential equations, focusing on Frobenius power-series solutions near regular singularities and their connection formulas. Central to our approach is the notion of a cline in the complex plane, which organizes singular points of the differential equations and remain invariant under Möbius transformations. Building on the cline structure we identified in black hole horizons, we carry out a systematic reduction and relocation of poles in the differential equation to obtain explicit representations of the solutions. We illustrate our approach by extracting the scalar perturbation solutions for the $7$-dimensional Myers-Perry black hole and deriving the static scalar tidal Love numbers. These results suggest that clines expose a Möbius-invariant order within black hole perturbations, rendering black hole perturbation problems remarkably tractable.

\end{abstract}
\newpage
\tableofcontents

\section{Introduction}

Heun differential equations appear across physics—from quantum mechanics and condensed matter systems to perturbations in curved spacetimes—encoding field dynamics through their singularities and connection structure. 
In black hole physics, both in four and higher dimensions, Heun equations are particularly relevant to the study of
quantum corrections to black hole entropy \cite{Arnaudo:2024bbd, Arnaudo:2025btb}, scattering processes and greybody factors \cite{Cvetic:1997uw, Cvetic:2009jn, Jorge:2014kra}, quasinormal modes \cite{Berti:2009kk, Fioravanti:2021dce, Hatsuda:2021gtn, Bianchi:2021mft, Noda:2022zgk, Lei:2023mqx, Silva:2025khf, Chen:2025sbz}, black hole jets \cite{Camilloni:2023wyn} and tidal Love numbers \cite{Kol:2011vg, Hui:2020xxx, Hui:2021vcv, Hui:2022vbh, Rodriguez:2023xjd, Perry:2023wmm, Charalambous:2023jgq, Perry:2024vwz, Glazer:2024eyi}. Beyond this context, similar techniques have found applications in holography, particularly in the study of renormalization group flows \cite{Caceres:2021fuw, Caceres:2022smh}, correlation functions \cite{Gutperle:2025bzr} and deformations of the boundary conformal field theory \cite{Karch:2025hof}. 

Over the last few years, the Heun connection formula program has achieved extraordinary success in describing black hole perturbations, including scattering processes \cite{Bonelli:2021uvf, Bonelli:2022ten}. This was first implemented to calculate Schwarzschild and Kerr quasinormal modes \cite{Aminov:2020yma}\footnote{We thank Alba Grassi for clarifying this point to us}. The method addresses the connection problem — which involves establishing the matrices that connect Frobenius series solutions around distinct singular points — for the local solutions of the four-point Heun equation, employing the so-called AGT correspondence introduced by Alday, Gaiotto, and Tachikawa \cite{Alday:2009aq}. A central aspect is the use of conformal blocks provided by this duality to construct explicit connection formulas between the local solutions of the Heun equation \cite{Bonelli:2022ten} and for Generalized Heun equations with five poles \cite{Arnaudo:2025kof}. This approach has also been generalized to the confluent limit of the Heun equation \cite{Bonelli:2021uvf, Bonelli:2022ten, Consoli:2022eey, Aminov:2023jve,  Jia:2024zes, Gutperle:2025bzr}. An alternative approach for deriving connection formulas is based on expressing them as ratios of Wronskians constructed from independent solutions of the differential equation. This method, grounded in the analytic continuation properties of linear differential equations, allows one to extract connection coefficients without explicitly matching local Frobenius series. As demonstrated by Hatsuda, Kimura, Noda, and Motohashi in the study of quasinormal modes~\cite{Hatsuda:2021gtn, Noda:2022zgk}, the Wronskian approach offers both analytical and numerical efficiency \cite{olver1997asymptotics, ince2012ordinary, slavjanov2000special}. Both of these methods should be linked, as they share the same underlying mathematical structures.

In Kerr black hole perturbation theory, the radial and angular scalar perturbation equations reduce to confluent Heun forms, yet exact analytic control remains complex and much of the underlying structure obscured \cite{Leaver:1986vnb, Suzuki:1998vy, Suzuki:1999nn, Sasaki:2003xr, Bautista:2023sdf}. Only in special limits does the Heun equation degenerate into a hypergeometric form — the simplest three points descendant of the Gauss equation \cite{Dias:2009ex, Bredberg:2009pv, Yang:2013uba, Castro:2013kea, Porfyriadis:2014fja, Compere:2017hsi, LeTiec:2020bos,  Castro:2021csm,  Charalambous:2021mea, Ivanov:2022qqt, Perry:2023wmm, Cvetic:2024dvn, Perry:2024vwz}. Well-known examples include the Legendre and Hermite equations, among others \cite{Bateman:100233}. This reduction enables exact solutions to the black hole perturbation problem, leading to the development of the hidden Kerr/CFT correspondence \cite{Guica:2008mu} and exact characterization of the static Love number \cite{LeTiec:2020bos, Charalambous:2021mea, Ivanov:2022qqt, Rodriguez:2023xjd, Perry:2023wmm, Perry:2024vwz}. It is worth noting that, beyond the Frobenius method, other prominent approaches for solving black hole perturbations include the technique introduced by Leaver \cite{Leaver:1986vnb} and subsequently extended by Mano, Suzuki, and Takasugi \cite{Suzuki:1998vy, Suzuki:1999nn, Sasaki:2003xr}.

Beyond four regular singular points, one encounters Generalized Heun equation \cite{ronveaux1995heun}. Although the Frobenius method can still be applied, and the overall solution approach is similar, the main challenge lies in connecting the local solutions and extending their domain within the solution space. To overcome this, physically motivated approximations are often employed to broaden the range of validity \cite{Jorge:2014kra}.

The goal of this paper is to show that the Generalized Heun equations arising in the study of black hole perturbations can be reduced to tractable singular four-point Heun-type problems. Our proposal exploits the \textit{Möbius covariance} of the analytic structure defined by the regular singular points in the complex plane. Specifically, we find that the singular points in the complex plane lie on circles and lines, referred to as \textit{clines}, which are themselves invariant under Moebius transformations \cite{hitchman2009geometry}. We employ these findings, together with specific properties of the background, to reduce our original Generalized Heun problem into a (4-point) Heun differential equation via a suitable coordinate transformation and field redefinitions. By applying appropriate boundary conditions, we obtain the local Heun functions, which are then related through connection formulas written as ratios of Wronskians, providing an exact and fully analytic solution to the equations.

For the sake of concreteness and with a specific application to the black hole perturbation problem in
mind, we formulate the solutions to the Klein-Gordon equation in the Myers–Perry black hole background to derive the tidal Love numbers. More precisely, we derive an exact solution to the scalar perturbation equation in seven dimensions that satisfies the appropriate boundary conditions for the tidal response. Then, considering the small-spin limit, we obtain the scalar Love numbers and find that they are nonzero. As a consistency check, taking the spinless limit reproduces the known Schwarzschild result \cite{Hui:2020xxx}. Unlike the four-dimensional case, scalar tidal Love numbers for higher-dimensional black holes generally do not vanish \cite{Charalambous:2023jgq, Rodriguez:2023xjd, Glazer:2024eyi, Gray:2024qys}. With the exception of the Schwarzschild–Tangherlini and five-dimensional Myers–Perry cases \cite{Hui:2020xxx, Rodriguez:2023xjd}, most black hole perturbation computations were performed under well-motivated (near-horizon) approximations. Remarkably, with our construction we demonstrate that the perturbations of higher-dimensional black holes can be made even more precise.

This paper is organized as follows. Section 2 introduces the key mathematical background used for the generalized Heun equation eigenfunction, the concepts of clines and their invariance under Mobius transformation. In Section 3, we present the implementation of this framework for the seven-dimensional Myers-Perry black hole background. We consider the metric of interest and describe its cline structure which is associated with the invariance of the Klein-Gordon equation under Möbius transformations. By applying suitable coordinate transformations and enforcing the boundary conditions necessary for computing the Love numbers, we obtain an exact solution to the perturbation equations. In Section 4, we explicitly calculate the scalar Love number for this system and establish its connection with previously known results. We conclude with a discussion in Section ~\ref{sec:Discuss}. In Appendix~\ref{sec:AppeWronsk}, we present the implementation of connection formulas using ratios of Wronskians, focusing specifically on the case of two consecutive regular singular points on the real line at \(z=0\) and \(z=1\), due to their relevance for the current discussion. A similar procedure can be applied to any pair of consecutive regular singular points. In Appendix~\ref{sec:MP5Love}, we provide the corresponding results for the five-dimensional Myers-Perry black hole, based on the calculations in \cite{Rodriguez:2023xjd}. Finally, in Appendix~\ref{sec:MathIden}, we review useful mathematical identities involving the Heun, Gamma, and hypergeometric functions.

\section{Clines, Generalized Heun Equation and Black Holes}

In this section, we introduce the concept of clines, which underlie the Mobius covariant structure of the generalized Heun differential equation. For concreteness, we focus on clines for higher $D$-dimensional black holes. However, the framework involving clines can be readily generalized to a wide range of physical problems, including black holes with cosmological horizons, such as the Kerr-AdS and BTZ geometries \cite{Guica:2014dfa, Dias:2019ery, Kajuri:2020bvi}.
We then review the general framework of connection formulas expressed in terms of ratios of Wronskians, which will later be employed in Section 3 to obtain explicit solutions. In addition, we outline the general procedure for computing scalar Love numbers. Together, these elements constitute the mathematical foundation underlying our proposed framework.

\subsection{Generalized Heun Equation}

Many black hole perturbation equations can be reduced to generalized Heun equations -- an extension of the classical Heun equation, describing a second-order linear ordinary differential equation (ODE) with more than three regular singular points. 
For instance, after separation of variables, the scalar perturbation equations for Kerr or Myers--Perry black holes lead to Heun-type equations in either the radial or angular sectors. 
We therefore establish the notation in this subsection and clarify the different types of reductions to hypergeometric form, as well as the corresponding solutions obtained via Frobenius expansions.

The generalized Heun equation \cite{ronveaux1995heun, NIST:DLMF, slavjanov2000special}, featuring 
$n + 1$ regular singularities, being $n$ finite singularities in addition to the singularity at infinity, can be written as
\begin{align}
\label{eqn:gen}
\frac{d^2y}{dz^2} + \left[ \sum_{i=1}^{n} \frac{\gamma_i}{z-z_i}\right] \frac{dy}{dz} + \left[ \sum_{i=1}^{n} \frac{q_i}{z-z_i}\right] y = 0,
\end{align}
where $z_i$ are the finite singular points, $\{0, 1-\gamma_i\}$ are the characteristic exponents (monodromy parameters) of the finite singularities $z_i$, $\{\alpha, \beta\}$ the characteristic exponents for the singular point at $\infty$ \footnote{The characteristic exponents $\rho$ are determined as the roots of the indicial equation associated with \eqref{eqn:gen},
\bea
\label{eqn:IndEq1}
\rho\,  (\rho-1) + b_i \, \rho + c_i =0\,,
\eea
\bea
b_i = \mathrm{Res}_{z=z_i} \sum_{i=1}^n \frac{\gamma_i}{z-z_i}\,,\qquad
c_i = \mathrm{Res}_{z=z_i}(z-z_i) \sum_{i=1}^n \frac{q_i}{z - z_i}\,,
\eea
associated to each finite regular singular point. For the singular point at infinity, the coefficients in Eq.~\eqref{eqn:IndEq1} are replaced as
\bea
b_i \rightarrow b_{\infty} = -\mathrm{Res}_{z=\infty} \sum_{i=1}^n \frac{\gamma_i}{z-z_i}\,,\qquad
c_i \rightarrow c_{\infty} = \mathrm{Res}_{z=\infty}z \sum_{i=1}^n \frac{q_i}{z - z_i}\,.
\eea}, and $q_i$ the accessory parameters. The accessory parameters $q_i$, finite regular singular points $z_i$ and $z=\infty$ characteristic exponents $\{\alpha, \beta\}$ 
satisfy the following relations
\bea
\label{eqn:genrel}
\sum_{i=1}^n q_i = 0\,, && \alpha\beta = \sum_{i=1}^n z_i q_i\,.
\eea  
The equation therefore has only $n-2$ independent accessory parameters.

For differential equations of the form \eqref{eqn:gen} with only regular singular points, the so-called Fuchsian relation, or balance condition on the residues of the connection, is satisfied.
\bea
\label{eqn:FuchsR1}
\alpha + \beta + 1 = \sum_{i=1}^n {\gamma}_i \,.
\eea
where $n$ refers to the number of finite regular singular points of the equation.

In this generalized Heun framework, the equation can have five, or more singularities. In the special case $n=3$ with only four regular singularities at $(z_i,\infty)$ with $i=1,2,3$, the generalized Heun equation \eqref{eqn:gen} reduces to the standard Heun equation. Further reduction, when $n=2$, \eqref{eqn:gen} yields the hypergeometric equation, which possesses only three regular singular points. The hypergeometric, Heun and generalized Heun equations, all satisfy \eqref{eqn:FuchsR1}. Irregular singular points appear in this context due to the coalescence of singularities. A summary of the main classes of Heun equations and their confluent cases is presented in Table~\ref{tab:heun_symmetries}.

\begin{table}[h!]
\centering
\small
\renewcommand{\arraystretch}{1.25}
\setlength{\tabcolsep}{3pt}
\begin{tabular}{|l|l|l|l|l|}
\hline
\textbf{Diff. Equation} &
\textbf{Singularities } \\
\hline

Hypergeometric (Gauss) &
3 regular \\ 
\hline

Heun (standard) &
4 regular \\ 
\hline

Generalized Heun &
\(n \ge 4\) regular\\
\hline

Confluent Heun (CHE) &
2 regular + 1 irregular (rank 1)\\ 
\hline

Biconfluent Heun (BCH) &
1 regular + 1 irregular (rank 2) \\ 
\hline

Double-confluent Heun (DCH) &
2 irregular (often at \(0,\infty\)) \\ 
\hline

Triconfluent Heun (TCH) &
1 irregular (rank 3)  \\ 
\hline
\end{tabular}
\caption{Nature of singularities for hypergeometric, Heun, and confluent Heun families.}
\label{tab:heun_symmetries}
\end{table}
Local solutions to the generalized equation can be obtained through Frobenius power-series expansions. 
If $z_0$ is a regular singular point, the solution can be expanded as
\begin{equation}
\label{eq:frobenius}
y(z) = (z - z_0)^{\rho} \sum_{n=0}^{\infty} a_n (z - z_0)^n ,
\end{equation}
where $\rho$ is determined from the corresponding indicial equation, with $a_0 \neq 0$.

The local solutions around each singular point can be connected using what is known as a \textit{connection formula}. Various techniques, such as monodromy methods \cite{Castro:2013lba,Castro:2013kea,CarneirodaCunha:2015hzd}, Wronskian methods Ref.~\cite{Hatsuda:2021gtn, Noda:2022zgk}, and \emph{AGT} approach \cite{Bonelli:2022ten,Bonelli:2021uvf}, have been developed for this purpose. In the following, we will focus on the Wronskian method, a summary of which is provided in {Appendix~\ref{sec:AppeWronsk}.


\subsection{Clines and Differential Equations}

 A \emph{cline} is a circle or a line in the complex plane along which singularities can be mapped using a Möbius (linear fractional) transformation. A classical result in geometry states that any three points in the complex plane lie on a circle or a line. Since there are only three points, any M\"obius transformation can map them to $(0, 1, \infty)$ in any chosen order. This M\"obius invariance expresses the projective symmetry of the Riemann sphere with three punctures. The symmetries of the hypergeometric differential equation (e.g. in BTZ and NHEK black holes perturbation \cite{Guica:2014dfa, Chen:2017ofv, Dias:2019ery, Kajuri:2020bvi, Castro:2021csm}) are a deep and elegant topic, connecting differential equations, projective geometry, and group theory.

When a fourth point is added, only the cross-ratio
\bea
\lambda = \frac{(z_4 - z_1)(z_3 - z_2)}{(z_4 - z_2)(z_3 - z_1)}
\eea
remains invariant under Möbius transformations. Thus, only an 
\( S_4 \) subgroup survives, corresponding to the permutations of the four singularities. Consequently, the singular points of the standard Heun equation can always be considered as lying on a cline and invariant under Möbius transformations. This allows the four singularities of Heun’s equation to be mapped onto the real axis or a circle, i.e., onto a cline where now the singularities are $(0,1,a,\infty)$. 
Given four distinct complex points $z_1, z_2, z_3, z_4$, define the Möbius map $f$ that sends
\bea
z_1 \mapsto 0, \quad z_2 \mapsto 1, \quad z_3 \mapsto \infty
\eea
by
\bea
f(z) = \frac{(z - z_1)(z_2 - z_3)}{(z - z_3)(z_2 - z_1)}.
\eea
The image of the fourth point, $f(z_4) = \lambda$, is real if and only if the four $z_i$ points lie on a cline. 
Thus, after an appropriate Möbius transformation, the singularities of the standard Heun equation can always be mapped to a single cline via Möbius transformations.

In the generalized Heun case, when there are more than four finite singularities, it is generally not possible for all of them to lie on a single circle or line. Hence, the geometric configuration of the singularities becomes more intricate  reflecting the increased geometric complexity of the generalized case \footnote{As noted earlier, generalized Heun equations satisfy the Fuchs relation \eqref{eqn:FuchsR1}. This condition, which arises from the presence of only regular singular points, places them within the so-called class of \emph{Fuchsian equations}. These equations can be divided into two main categories—\textit{reducible} and \textit{irreducible}—based on their covariance under \textit{field redefinitions} of the dependent variable. In the mathematical literature, such transformations are referred to as \textit{s-homotopic transformations}. For a detailed formal discussion, see Slavyanov and Lay \cite{slavjanov2000special}.}.. One can still use Möbius transformations to simplify three of the singular points (e.g., mapping them to $0$, $1$, $\infty$), but the remaining singularities may lie off a single cline. We will now argue that the singularities in black hole perturbation lie on clines, and hence are Moebius invariant. We will later employ this fact to unravel the Myers Perry in 5D and with a precise example in 7D show that the reduce the poles to become a Heun equation.

We emphasize in this section that

\noindent \textit{the geometric arrangement of the singular points of an ODE lying on a cline
remains invariant under Möbius transformations.}

This invariance provides a powerful tool for analyzing the symmetry and structure of the differential equation, and it underlies many of the simplifications used in the study of Heun and hypergeometric-type of differential equations.

We now show that the singularities arising in black hole perturbations equations lie on clines and are therefore are invariant under Möbius transformations. We detail how this property was exploited to analyze perturbations of the five-dimensional Myers-Perry black hole \cite{Cvetic:1997uw, Rodriguez:2023xjd} and exploit this structure for further applications. As a concrete example in seven-dimensional Myers-Perry black holes, we show how the pole structure of the (radial) scalar perturbations differential equation, governed by generalized Heun equations, can be systematically reduced to yield equations of Heun type.

\subsection{Clines and Black Holes}
We begin by analyzing the analytic structure of black hole perturbations propagating in the background of interest. To illustrate our findings, we focus on the Myers-Perry solution in odd dimensions with all angular momentum parameters set equal. The horizons are determined by solving the equation
\bea
\label{eqn:HorCon1}
g^{rr}=0\,.
\eea
In the present case, we consider a $D = 2N + 3$ dimensional metric, where $N + 1$ represents the number of independent rotation planes. Equation \eqref{eqn:HorCon1} yields the condition for the horizon,
\bea
\label{eqn:HorCon2}
\Delta (r) \equiv r^{2N+2} - r_{s}^{2N}r^{2} + r_{s}^{2N}a^2 = 0
\eea
The zeros of Eq.~\eqref{eqn:HorCon2} correspond to the horizon radii of the black hole and include the event horizon $r_{+}$ and the Cauchy horizon $r_{-}$, both of which are real in this case, while the remaining singular points are, in general, complex.  We examine Eq.~\eqref{eqn:HorCon2} for the cases $D = \{9, 11, 13\}$. The corresponding configurations of singular points in the complex plane are shown in Fig.~\ref{fig:HorOddMP}.
\begin{figure}[!ht]

\centering
\begin{subfigure}{.32\textwidth}
    \centering
    \includegraphics[width=\linewidth]{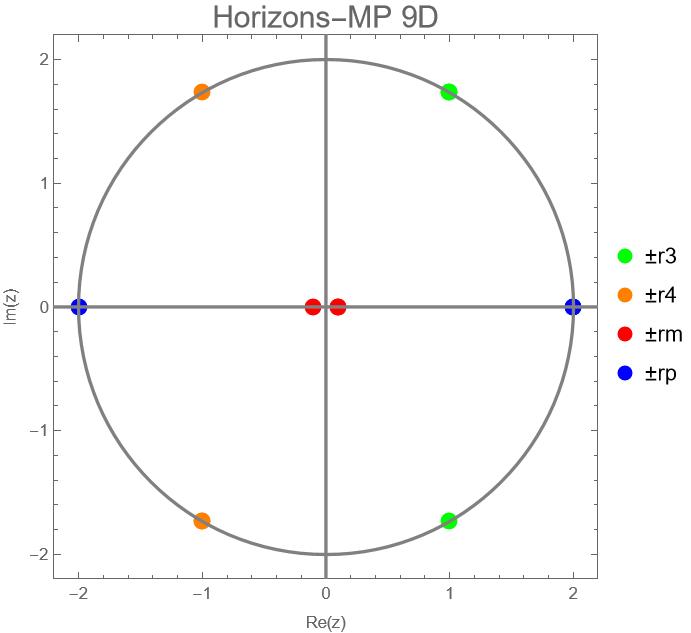}
\end{subfigure}
\hfill
\begin{subfigure}{.32\textwidth}
    \centering
    \includegraphics[width=\linewidth]{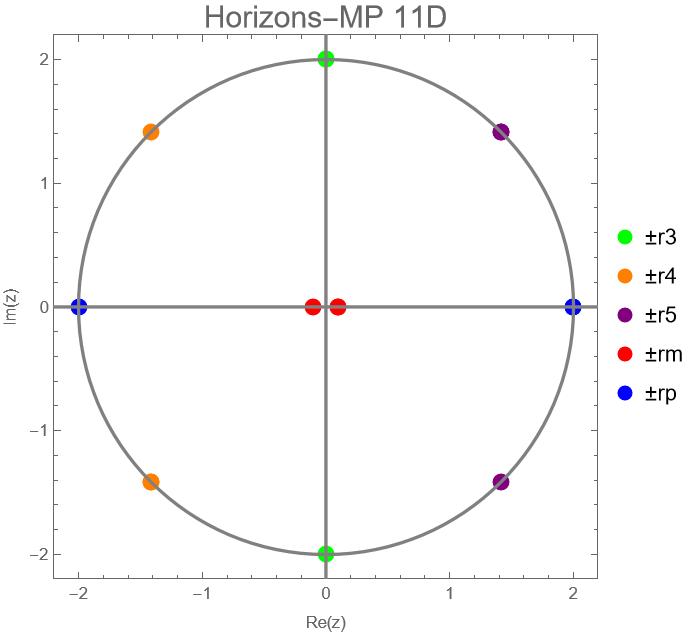}
\end{subfigure}
\hfill
\begin{subfigure}{.32\textwidth}
    \centering
    \includegraphics[width=\linewidth]{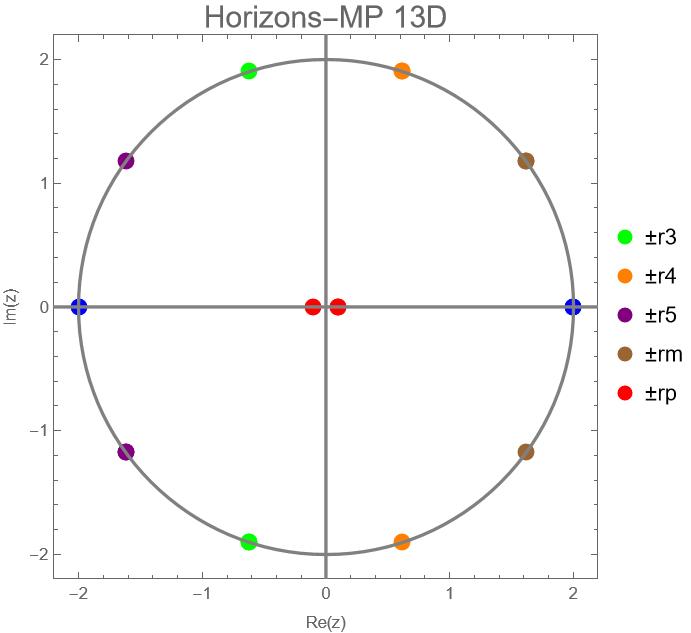}
\end{subfigure}
\caption{Complex-plane representation of the black hole horizons for Myers--Perry solutions in $D=9,11,13$ dimensions with all angular momentum parameters set equal. We used as mass and angular momentum parameters $\{r_s, a\} = \{2, 0.1\}$, respectively. The seven-dimensional case will be discussed separately in Section~\ref{sec:3}. Double poles are indicated by the same color, and the point at infinity is not shown in the plots.
}
\label{fig:HorOddMP}
\end{figure}

All horizons are found in a cline (on either a line or a circle) in the complex plane. Their geometric location can be expressed mathematically as
\begin{align}
\label{eqn:ClineG}
cz\bar{z} + \alpha z + \bar{\alpha}\bar{z} + d =0\,. 
\end{align}
Here $\alpha, z, \bar{z} \in \mathbb{C}$ and $c, d \in \mathbb{R}$.
As shown in Fig.~\ref{fig:HorOddMP}, the singular points are located on the real and imaginary axes, as well as on circles centered at the origin. For points lying on lines, the cline equation \eqref{eqn:ClineG} reduces to
\begin{align}
\label{eqn:ClineL}
\alpha z + \bar{\alpha}\bar{z} + d =0\,.
\end{align}
In any case, clines \eqref{eqn:ClineG}\eqref{eqn:ClineL} are invariant under Mobius transformations \cite{needham2023visual, conway2012functions, hitchman2009geometry, lang2013complex},
\begin{align}
\label{eqn:Mobius1}
T(z)=\frac{az+b}{cz+d}\,,
\end{align}
where $\alpha, z, \bar{z} \in \mathbb{C}$ and $c, d \in \mathbb{R}$. 

 Specifically it seems that the invariance is discrete. As with clines, the differential equations derived for this background are Mobius invariant.
Coincidentally, all black hole horizons, along with infinity, correspond to the regular singular points of the differential equation governing the scalar black hole perturbations, namely the Klein--Gordon ($s=0$ Teukolsky) equation as we further detail in Section \ref{sec:3}.

\begin{figure}[!ht]
\centering
\begin{subfigure}{.32\textwidth}
    \centering
    \includegraphics[width=\linewidth]{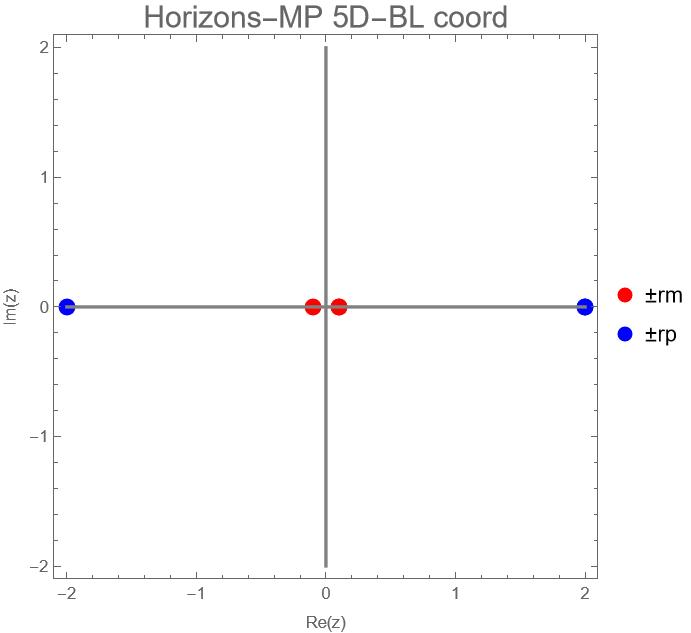}
\end{subfigure}
\begin{subfigure}{.32\textwidth}
    \centering
    \includegraphics[width=\linewidth]{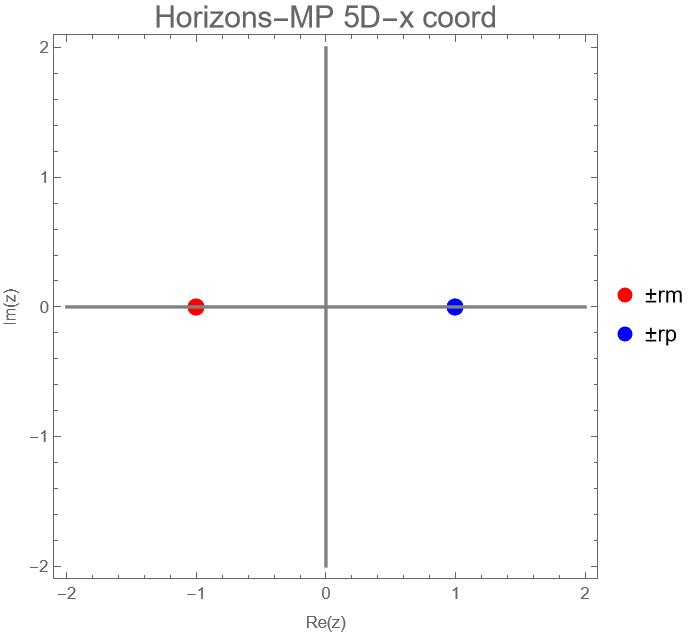}
\end{subfigure}
\begin{subfigure}{.32\textwidth}
    \centering
    \includegraphics[width=\linewidth]{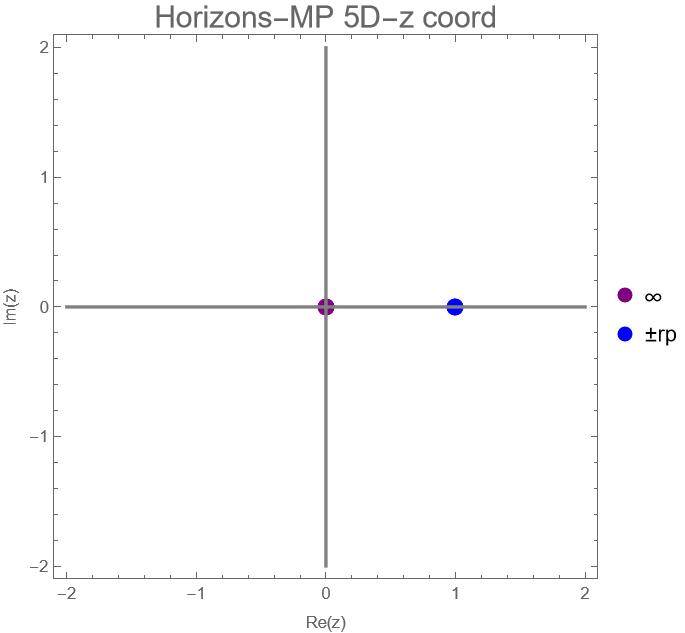}
\end{subfigure}
\caption{Complex-plane representation of the black hole horizons for the five-dimensional Myers–Perry solution belonging to a cline. After applying specific Möbius and diffeomorphism transformations, the singular points can be rearranged along the real axis in the complex plane. This procedure effectively reduces the corresponding differential equation to a hypergeometric form, rendering the problem analytically solvable. This approach was first illustrated in the work of Cvetič et. al \cite{Cvetic:1997uw} and further employed for the computation of Love numbers in \cite{Rodriguez:2023xjd}.} 
\label{fig:MPCoordCline}
\end{figure}

Focusing on clines and Möbius transformations, let us reassess the five-dimensional Myers Perry black hole case. This symmetry was first made manifest by Myers and Perry~\cite{Myers:1986un}, and later elucidated by Cveti\v{c} and Larsen~\cite{Cvetic:1997uw} through a combination of a coordinate (diffeomorphism) transformation and a nonlinear mapping involving $r^2$, composed with a Möbius transformation to simplify the analytic structure. Our nonlinear map acts on our radial coordinate $r$,
\bea
\label{eqn:CveLar}
x=\frac{2 r^2 - (r_+^2 + r_-^2)}{(r_+^2 - r_-^2)}\,,
\eea
with the poles $r=\{\pm r_-, \pm r_+, \infty\}$ being mapped to $x=\{-1, 1, \infty\}$, respectively.
Next we introduce a Mobius transformation \eqref{eqn:Mobius1} of the form
\bea
\label{eqn:MobMP5}
z = \frac{2}{x+1}\,,
\eea
so the poles $x=\{-1, 1, \infty\}$ get to $z=\{\infty, 1, 0\}$. Together \eqref{eqn:MobMP5} and \eqref{eqn:CveLar} lead us to propose a rational map of order 2
\bea
z=\frac{r_+^2 - r_-^2}{r^2 - r_-^2}\,.
\eea
In many black hole metrics (e.g., Myers--Perry or Kerr), the equations naturally depend on $r^2$ rather than on $r$ itself. This arises because the metric functions and singular structures are typically even in $r$, with horizons located at the zeros of such even functions. Consequently, it is convenient to introduce a new variable
$x = r^2,$ which renders the differential equation rational in $x$. This mapping is nonlinear and therefore not a Möbius transformation; it preserves conformality but does not preserve clines globally. However, a subsequent Möbius transformation is applied to the variable $x = r^2$, the cline structure is preserved in the $x$-plane, although it is not preserved in the original $r$-plane. In this way, one can map clines to clines in the solution space while preserving the Fuchsian structure of our generalized Heun equation. This has been illustrated for the five dimensional case (see Fig.~\ref{fig:MPCoordCline}). We argue that this approach can be generalized to other Myers--Perry black hole perturbations. More generally, it applies whenever the singularities of the differential equations define clines.

A preliminary analysis of the general Myers--Perry case (in all $D$ dimensions, including even dimensions) suggests that the structure of the horizons in the complex plane is geometrically organized along clines (see Fig.~\ref{fig:HorDifSpin}). Remarkably, this pattern persists for all values of the spin, which need not be equal. For clarity, we illustrate the analysis in odd dimensions; nevertheless, we argue that this Moebius hidden symmetry among the horizons is a general feature present in all black hole solutions that we have analyzed so far.

\begin{figure}[H]

\centering
\begin{subfigure}{.32\textwidth}
    \centering
    \includegraphics[width=\linewidth]{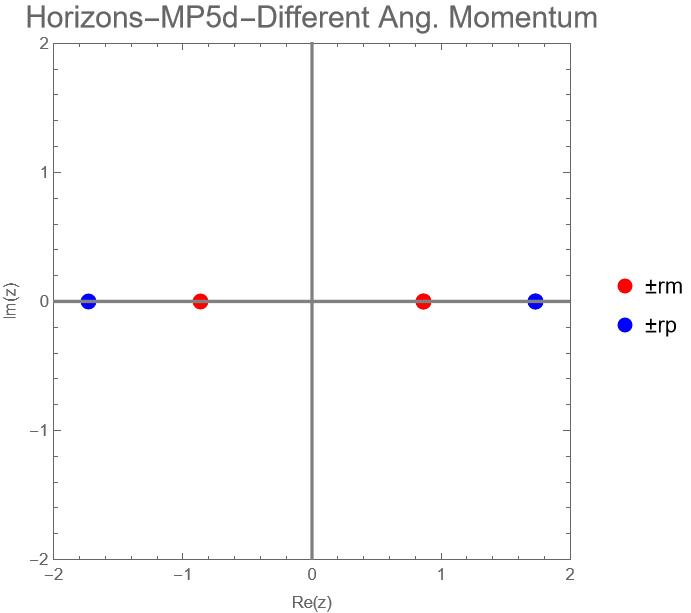}
\end{subfigure}
\begin{subfigure}{.32\textwidth}
    \centering
    \includegraphics[width=\linewidth]{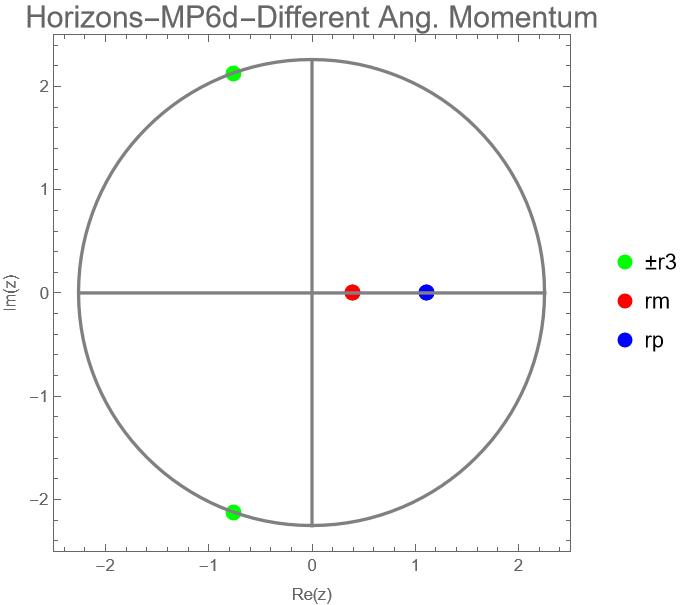}
\end{subfigure}
\begin{subfigure}{.32\textwidth}
    \centering
    \includegraphics[width=\linewidth]{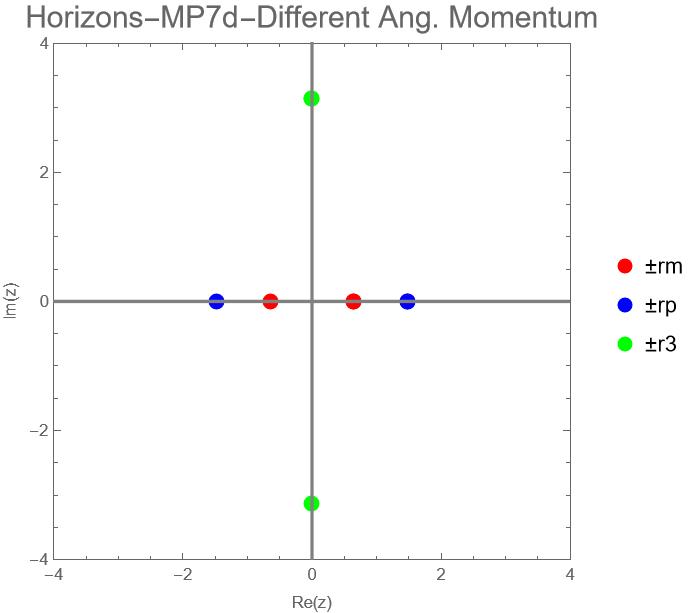}
\end{subfigure}
\hfill

\caption{Complex-plane representation of the black hole horizons for the five-, six-, and seven-dimensional Myers--Perry solutions \cite{Myers:2011yc}. For the five-dimensional case, the mass and angular momentum parameters are $\{\mu,a,b\} = \{1, 1.5, 7\}$ respectively. For the solution in 6 dimensions, we have $\{\mu, a, b\}=\{7, 1,1.5\}$. In seven dimensions we used $\{\mu, a, b, c\} = \{40, 1, 1.5, 2\}$. The  resulting horizons are found to lie along clines.}
\label{fig:HorDifSpin}
\end{figure}
\section{Perturbations in D-dimensional Myers–Perry black hole}
\label{sec:3}

To illustrate the applicability of the geometric structure underlying the singularities of differential equations, in this section we present computations for scalar perturbations in higher-dimensional Myers–Perry black holes, specifically focusing on the seven-dimensional case. As we argued earlier, the reduction of the differential equation to a simpler form is possible because the singularities of the associated perturbation equation lie along clines in the complex plane. By applying appropriate Möbius transformations, the seven singular points of the radial differential equation can be mapped to four, thereby simplifying the analytic structure of the equation—from a generalized Heun form to the standard Heun equation. We then obtain an exact solution for the massless scalar perturbation equation. 


\subsection{Myers Perry Black Holes}
\label{Sec:Intro}

Myers and Perry \cite{Myers:1986un} extended the Kerr solution to higher-dimensional spacetimes. In odd dimensions ($D = 2N + 3$), the solution describes a spheroidal black hole with $N + 1$ independent rotational planes. The geometry is specified by a mass-radius parameter $r_s$ and $\frac{D-1}{2}$ angular momentum parameters $a_i$. In the special case where all angular momenta are equal ($a_i = a$), the metric can be written in a form with enhanced symmetry. 
 In general, the Myers-Perry metric is isometric under the action of the group
\bea
\label{eqn:IsoMP1}
G=\mathbb{R}\times U(1)^N\,,
\eea
but when the angular momenta are set equal, its symmetry group enhances to
\bea
\label{eqn:IsoMP2}
G=\mathbb{R}\times U(N)\,.
\eea
The orbits of this group are hypersurfaces of constant $r$, in other words, squashed $S^{2N-1}$ spheres. This characterizes the cohomogeneity-1 geometry.

Following the conventions of \cite{Dias:2010eu, Jorge:2014kra}, the line element takes the following form:
\begin{equation}
\label{eqn:MPodd1}
ds^2 = -f(r)^2 dt^2 + g(r)^2 dr^2 + h(r)^2 \left[d\psi + A_a dx^a - \Omega(r) dt\right]^2 + r^2 \hat{g}_{ab} dx^a dx^b, 
\end{equation}
where
\begin{align}
\label{eqn:MPodd2}
g(r)^2 &= \left(1 - \frac{r_s^{2N}}{r^{2N} }+ \frac{r_s^{2N} a^2}{r^{2N+2} }\right)^{-1}, \qquad h(r)^2 = r^2 \left(1 + \frac{r_s^{2N}a^2}{r^{2N+2} }\right), \\
f(r) &=\frac{ r}{ g(r) h(r)}, \qquad \Omega(r)= \frac{r_s^{2N}a}{r^{2N} h(r)^2}.
\end{align}
In the above equations, $\hat{g}_{ab}$ denotes the Fubini–Study metric on the complex projective space $\mathbb{CP}^N$, and $A = A_a dx^a$ corresponds to its Kähler potential. Explicit expressions for these quantities can be obtained recursively in $N$ (see, e.g., Ref.~\cite{Dias:2010eu}). The coordinate $\psi$ parametrizes the $S^1$ fiber and has period $2\pi$.

Here the radial coordinate is related to the standard Boyer-Lindquist radial coordinate as defined in \cite{MyersPerry, Myers:1986un}, through the mapping $r^2\rightarrow r^2 +a^2$. The spacetime metric satisfies $R_{\mu\nu} = 0$ and the solution is asymptotically flat. Moreover, like its Kerr counterpart, Myers Perry black hole solutions exhibit an extremality bound given by
\bea
\label{eqn:MPext}
\Big(\frac{a}{r_+}\Big)^2 \leq \Big(\frac{a_{ext}}{r_+}\Big)^2 = N\,, && \Big(\frac{a}{r_s}\Big)^2 \leq \Big(\frac{a_{ext}}{r_s}\Big)^2 = \frac{N}{(N+1)^{(N+1)/N}}\,. 
\eea
Distinct from the Kerr case, this higher dimensional black hole solution exhibit ultraspinning behavior when the following constraint is imposed
\bea
\label{eqn:MPultra}
\Big(\frac{a}{r_+}\Big)^2 > \Big(\frac{a_1}{r_+}\Big)^2 \equiv \frac{1}{2}\,, && \Big(\frac{a}{r_s}\Big)^2 > \Big(\frac{a_1}{r_s}\Big)^2 \equiv \frac{1}{2^{(N+1)/N}}\,,
\eea
which eventually would lead to an instability \cite{Bhattacharyya:2015dva}.

\subsection{Scalar Perturbations}
\label{Sec:methods}

In this section, we focus on the massless scalar perturbation equation governed by the Klein-Gordon equation in curved spacetime
\bea
\label{eqn:KG1}
\frac{1}{\sqrt{-g}}{\partial}_{\mu}\Big[\sqrt{-g}g^{\mu\nu}{\partial}_{\nu}\Phi\Big]=0\,.
\eea
The wave equation admits separation of variables under an anstaz
\bea
\label{eqn:MPansatz}
\Phi (t, r, \psi, x^a) = e^{-i \omega t + i m \psi} R(r) \mathbb{Y}_{\ell m}(x^a)\, ,
\eea
Here $\mathbb{Y}$ represents an eigenfunction of the {\it charged} scalar Laplacian in $CP^N$ and its eigenvalues are given by $ K_{\ell,m}$ \cite{Kunduri:2006qa, Jorge:2014kra},
\be
\begin{aligned}
\label{eqn:ChargedLap}
- \mathcal{D}^2 \mathbb{Y}_{\ell m} &\equiv -\hat{g}^{ab}\left(D_a - i m B_a\right) \left(D_b - i m B_b\right)\mathbb{Y}_{\ell m} 
\\
&= \left[\ell(\ell +2N)  - m^2\right]\mathbb{Y}_{\ell m} \, .
\end{aligned}
\ee
where $\mathcal{D}$ denotes the gauge-covariant derivative on $\mathbb{CP}^N$, and the eigenvalues are given by $K_{\ell,m} = \ell(\ell + 2N) - m^2$, with $\ell = 2k + |m|$ and $k = 0, 1, 2, \ldots$. Note that for a given $\ell$, the spacing between consecutive values of $m$ is always even.

The radial equation can be now cast into the following form (see e.g. \cite{Jorge:2014kra})
\begin{align}
\label{eqn:MPKGequalspins}
\partial_r \left[\frac{r^{2N+1}}{g(r)^2} \partial_r R(r) \right] +
 \Bigg[&g(r)^2\frac{r^2 \left(r_s^{2N} a^2 +r^{2N+2}\right)^2 (\omega -m \Omega (r))^2}{r^{2N+2}}
 \\\nonumber
&- r_s^{2N}a^2  \Big(\ell( \ell +2 N)-m^2\Big)-\ell (\ell+2N) r^{2 N+2}\Bigg] \frac{R(r)}{r\,h(r)^2}=0 
\end{align}
By inspecting the structure of this function we can give a more suitable expression 
\bea
\label{eqn:BlackFunc}
g(r)^{-2}&=& \frac{\prod_{i=1}^{N+1}(r^2-r_i^2)}{r^{2N+2}}=\frac{\Delta_h(r)}{r^{2N+2}}\,,
\eea
where
\bea
\label{eqn:DeltaFunc}
\Delta_h(r)=(r^2-r_+^2)(r^2-r_-^2)\prod_{i=1}^{N-1}(r^2-r_i^2)\,.
\eea
The differential equation \eqref{eqn:MPKGequalspins} possesses finite $2N+2$ regular singular points, considently at the location of the black hole's horizons $r_h$ defined by $\Delta_h(r_h)=0$, along with an irregular singular point at $r = \infty$.
Here, we focus on the \textit{static} limit $(\omega = 0)$ of the equation, in which the point $r \rightarrow \infty$ becomes a regular singularity. In this limit, Eq.~\eqref{eqn:MPKGequalspins} reduces to the \textit{generalized Heun} form, featuring $2N+3$ regular singular points, including $r \rightarrow \infty$.

\bea
\label{eqn:KGMPStatic2}
{\partial}_{r}\Big[\frac{{\Delta}_h}{r}{\partial}_{r}R\Big]+\Big[\frac{a^2 m^2 r_s^{4N} r^2}{{\Delta}_h}-r_s^{2N} a^2 (\ell(\ell + 2N)-m^2) - \ell (\ell + 2N)r^{2N+2}\Big]\frac{R}{r\,h^2(r)}=0 .
\eea

The singular points of this equation are part of the cline, as illustrated in Fig. \ref{fig:HorOddMP}.
For the $D=5$ ($N=1$) solution, eq. \eqref{eqn:KGMPStatic2} reduces to the case studied in \cite{Cvetic:1997uw,Rodriguez:2023xjd}, when the angular momentum parameters are set equal, $a=b$. 

\subsection{The D=7 Myers-Perry}

We focus on the seven-dimensional ($D=7$, i.e.~$N=2$) case corresponding to the Myers--Perry black hole. In this case, Eq.~\eqref{eqn:KGMPStatic2} simplifies to
\bea
\label{eqn:KGMPStatic7d}
{\partial}_{r} \Big[\frac{\Delta_h}{r}{\partial}_{r}R\Big]+ \Big[-\ell(\ell+4)(r^6 + r_s^4 a^2) + m^2 r_s^4 a^2 + \frac{m^2 a^2 r_s^8 r^2}{{\Delta}_h}\Big]\frac{r^3 R(r)}{r^6 + r_s^4 a^2}=0\,,
\eea
The seven regular singular points of the equation are located at $r = \{\pm r_i, \infty\}$ with $i = +, -, 3$, where each $r_i$ is defined as a root of $\Delta_h(r_i) = 0$. Among them, $\pm r_+$ and $\pm r_-$ are real, while $\pm r_3$ are complex. Their squares satisfy the hierarchy $r_3^2 < 0 < r_-^2 < r_+^2$, since $r_3$ is purely imaginary. Similar to other higher dimensional cases discussed above (see Fig.~\ref{fig:HorOddMP}), we find that the regular singular points of the Klein--Gordon equation \eqref{eqn:KG1} in the seven-dimensional case exhibit a comparable cline structure. These poles align along distinct clines in the complex plane, as illustrated in Fig.~\ref{fig:HorBL7D}.

\begin{figure}[H]
\centering
\includegraphics[scale=0.4]{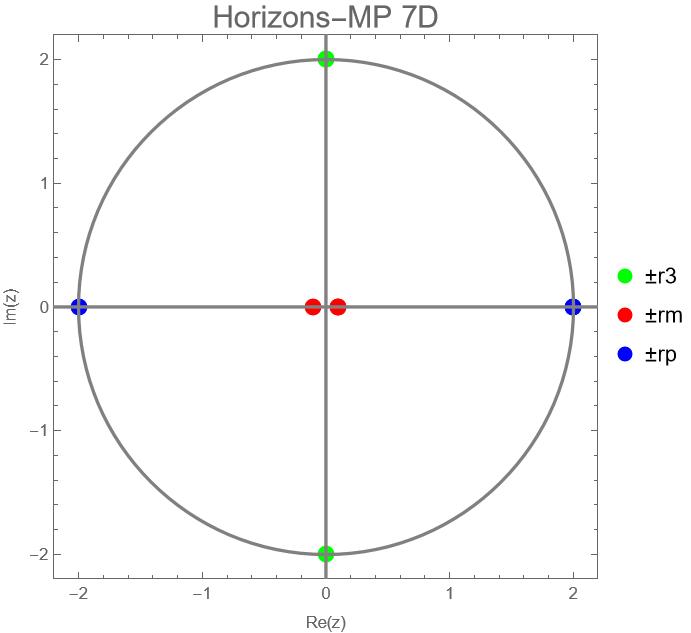}
\caption{Regular singular points for the scalar wave equation in Boyer-Lindquist coordinates.}
\label{fig:HorBL7D}
\end{figure}
In particular \eqref{eqn:DeltaFunc} takes the form
\bea
\label{eqn:Delta7D}
{\Delta}_h (r) = (r^2 - r^{2}_{+})(r^{2} - r^{2}_{-})(r^{2} - r^{2}_{3})\,.
\eea
For computational purposes we consider the next group of identities
\bea
r^{2}_{+} + r^{2}_{-} + r^{2}_{3} = 0\,,\\
r^{2}_{+}r^{2}_{-} + r^{2}_{+}r^{2}_{3} + r^{2}_{-}r^{2}_{3} + r^{4}_{s} = 0\,,\\
r^{2}_{+}r^{2}_{-}r^{2}_{3} + a^{2}r^{4}_{s} = 0\,,
\eea
rewriting eq. \eqref{eqn:KGMPStatic7d} as
\begin{align}
\label{eqn:KGMPStatic7d2}
{\partial}_{r}\Big[\frac{{\Delta}_h}{r}{\partial}_{r}R\Big]-\Big[\ell(\ell+4)\,r^{3} + \frac{m^{2}r^{2}_{+}r^{2}_{-}r^{2}_{3}r^{3}}{(r^{6}-r^{2}_{+}r^{2}_{-}r^{2}_{3})} - \frac{m^{2}r^{2}_{+}r^{2}_{-}r^{2}_{3}(r^{2}_{+}r^{2}_{-} + r^{2}_{+}r^{2}_{3} + r^{2}_{-}r^{2}_{3})r^{5}}{(r^{6}-r^{2}_{+}r^{2}_{-}r^{2}_{3}){\Delta}_h}\Big]R=0\,,
\end{align}
As a consistency check, we find that in the Schwarzschild limit ($r_{+}=r_{s}\,, r_{3}=ir_{s}\,, r_{-}=0$), the previous expression reduces to:
\begin{align}
\label{eqn:SchStatic7d}
{\partial}^{2}_{r}R(r)+\Bigg[\frac{5r^{4}-r^{4}_{s}}{r^{5}-r^{4}_{s}r}\Bigg]{\partial}_{r}R(r)-\Bigg[\frac{\ell(\ell+4)r^{2}}{r^{4}-r^{4}_{s}}\Bigg]R(r)=0\,,
\end{align}
coinciding with the result obtained in Refs.\cite{Hui:2020xxx, Glazer:2024eyi}.

Having identified the singularities of the differential equation, we now note that Eq.~\eqref{eqn:SchStatic7d} can be written in the form of a generalized Heun equation. While equations of this type have been widely analyzed in the literature, deriving explicit analytic solutions continues to be a challenging task.
\subsection{From Generalized Heun to Heun via Diffeomorphisms}

Our original motivation for analyzing this equation was to generalize the definition of the Love number, which requires solving perturbation equations involving generalized Heun functions. Having now identified an additional geometric structure—clines—associated with diffeomorphisms induced by Möbius transformations among three points, we proceed in this section to show how the equation can be reduced to a standard Heun form, rendering the problem more tractable.

We employ an affine transformation— a modified version of \eqref{eqn:CveLar} \footnote{This is as subset of  a full Möbius (fractional linear) transformation \refeq{eqn:Mobius1} with $c=0$, because the denominator is constant.} —whose analytical form is given by
\bea
\label{eqn:CveLarMod}
x=\frac{2 r^2 - (r_+^2 + r_-^2)}{r_+^2 - r_-^2}\,.
\eea
This is a fractional affine mapping of the variable $r^2$
 that sends the two horizon locations  to canonical symmetric point $(r_-^2,r_+^2) \rightarrow (-1,+1)$. This mapping was previously used to derive the scalar Love numbers in the five-dimensional Myers–Perry case \cite{Rodriguez:2023xjd}. Since this transformation preserves the cline structure inherent to \eqref{eqn:KGMPStatic7d2}, it can subsequently be composed with a full Möbius transformation
\bea
\label{eqn:Mobius2}
z=\frac{x-1}{x+1}\,,
\eea
and obtain
\bea
\label{eqn:RatMap}
z=\frac{r^{2}-r^{2}_{+}}{r^{2}-r^{2}_{-}}\,.
\eea
The transformation \eqref{eqn:RatMap}, together with the field redefinition
\begin{align}
\label{eqn:FieldRed}
R(z)=z^{\rho}(z-1)^{\sigma}(z-t)^{\tau}u(z)\,.
\end{align}
casts \eqref{eqn:KGMPStatic7d2} into the canonical form of the Heun equation
\begin{align}
\label{eqn:HeunCan1}
{\partial}^{2}_{z}u(z)+\Big[\frac{\gamma}{z} + \frac{\delta}{z-1} +\frac{\epsilon}{z-t}\Big]{\partial}_{z}u(z)+\Big[\frac{\alpha\beta z-q}{z(z-1)(z-t)}\Big]u(z)=0\,.
\end{align}
This is a well established transformation procedure in mathematics, extensively studied. For an in-depth review on the subject check \cite{ronveaux1995heun, maier2007192}.
Here the poles have been mapped so that $\pm r_- \rightarrow \infty, \pm r_+ \rightarrow 0, \pm r_3 \rightarrow t$ and $\infty \rightarrow 1$, where
\bea
t=\frac{r^{2}_{3}-r^{2}_{+}}{r^{2}_{3}-r^{2}_{-}}\,.
\eea
With this reshuffling, the points are arranged such that \( 0 < 1 < t < \infty \).

The field redefinition parameters are
\bea
\rho &=& - \frac{m r^{2}_{+}r_{-}r_{3}}{2(r^{2}_{3}-r^{2}_{+})(r^{2}_{+}-r^{2}_{-})}\,,\\
\sigma &=& -\frac{\ell}{2}\,,\\
\tau &=& -\frac{m r^{2}_{3}r_{+}r_{-}}{2(r^{2}_{3}-r^{2}_{+})(r^{2}_{3}-r^{2}_{-})}\,.
\eea
The characteristic exponent parameters for each regular singular point are
\begin{align}
\gamma = 1+2\rho\,, && \delta = -1+2\sigma\,, && \epsilon = 1+2\tau\,,
\end{align}
\begin{align}
\alpha = \rho + \sigma + \tau  + \frac{m r_{+}r_{3}r_{-}^{2}}{2(r^{2}_{+}-r^{2}_{-})(r^{2}_{3}-r^{2}_{-})}\,, && \beta = \rho + \sigma + \tau  - \frac{m r_{+}r_{3}r_{-}^{2}}{2(r^{2}_{+}-r^{2}_{-})(r^{2}_{3}-r^{2}_{-})}\,,
\end{align}
and the accessory parameter
\begin{multline}
q = A_1 \ell^2 + A_2 \ell + A_3 m^2 + A_4 \rho^2 + A_5\rho\sigma + A_6\rho\tau + A_7\rho + A_8\sigma +\tau\,,
\end{multline}
where
\bea
A_1 &=& \frac{r^{2}_{+}}{4r^{2}_{+}+8r^{2}_{-}}\,,\\
A_2 &=& \frac{r^{2}_{+}}{r^{2}_{+}+2r^{2}_{-}}\,,\\
A_3 &=& \frac{r^{2}_{+}r^{2}_{-}(r^{8}_{-}+8r^{6}_{-}r^{2}_{+}+17r^{4}_{-}r^{4}_{+}+13r^{2}_{-}r^{6}_{+}+3r^{8}_{+})}{4(r^{2}_{+}-r^{2}_{-})^{2}(r^{2}_{+}+2r^{2}_{-})(r^{6}_{-}+3r^{2}_{+}r^{4}_{-}+3r^{4}_{+}r^{2}_{-}+2r^{6}_{+})}\,,\\
A_4 &=& \frac{3(3r^{6}_{-}+8r^{4}_{-}r^{2}_{+}+8r^{2}_{-}r^{4}_{+}+2r^{6}_{+})}{2r^{6}_{-}+3r^{4}_{-}r^{2}_{+}+3r^{2}_{-}r^{4}_{+}+r^{6}_{+}}\,,\\
A_5 &=& \frac{(2r^{6}_{-}+6r^{4}_{-}r^{2}_{+}+6r^{2}_{-}r^{4}_{+}+4r^{6}_{+})}{2r^{6}_{-}+3r^{4}_{-}r^{2}_{+}+3r^{2}_{-}r^{4}_{+}+r^{6}_{+}}\,,\\
A_6 &=& \frac{(4r^{6}_{-}+6r^{4}_{-}r^{2}_{+}+6r^{2}_{-}r^{4}_{+}+2r^{6}_{+})}{2r^{6}_{-}+3r^{4}_{-}r^{2}_{+}+3r^{2}_{-}r^{4}_{+}+r^{6}_{+}}\,,\\
A_7 &=& \frac{(r^{2}_{-}-r^{2}_{+})}{2r^{2}_{-}+r^{2}_{+}}\,,\\ 
A_8 &=& \frac{(r^{2}_{-}+2r^{2}_{+})}{2r^{2}_{-}+r^{2}_{+}}\,.
\eea
The original cline structure is preserved, with all clines of the \textit{generalized Heun} equation mapped onto the real axis. More precisely, all singular points are mapped to the positive real line in the complex plane. This is illustrated in Fig.~\ref{fig:HorZ7D}.
\begin{figure}[h!]
\centering
\includegraphics[scale=0.4]{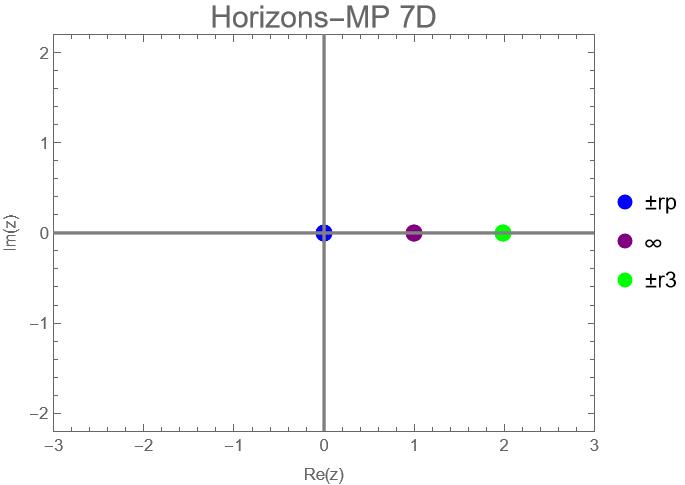}
\caption{Regular singular points of the scalar wave equation on z coordinates.}
\label{fig:HorZ7D}
\end{figure}

We can now determine the general form of the solutions to the Heun equation~\eqref{eqn:HeunCan1} by expanding around the regular singular points corresponding to the event horizon and spatial infinity. In doing so, we employ the Wronskian method, a powerful analytical technique that systematically relates local Frobenius solutions defined about distinct singular points. Following the formalism developed in Refs.~\cite{Hatsuda:2021gtn, Noda:2022zgk} (see also Appendix~\ref{sec:AppeWronsk} for detailed derivations), we first identify the local solutions as Heun local functions, \( Hl(z) \), around the event horizon \( z = 0 \) (which corresponds to the radial coordinate \( r = r_+ \))
\bea
&&u_{01} = Hl(t\,,q\,,\alpha\,,\beta\,,\gamma\,,\delta\,,z)\label{eqn:HorMP71}\,,\\
&&u_{02} = z^{1-\gamma} Hl(t,(t\delta + \epsilon)(1-\gamma)+q;\alpha+1-\gamma,\beta + 1 -\gamma,2-\gamma,\delta;z)\label{eqn:HorMP72}\,,
\eea
Further details on the Heun local function \( Hl(z) \) can be found in Appendix~\ref{sec:MathIden}.
The more general local solution to \eqref{eqn:KGMPStatic7d2} around the horizon is a linear combination of both
\bea
\label{eqn:HorMP73}
R(z)= z^{\rho}(z-1)^{\sigma}(z-t)^{\tau}\Big[B_1 u_{01}(z)+ B_2 u_{02}(z)\Big]
\eea
with constants $B_1$, $B_2$.

Similarly, the local solution around the sigular point at asymptotic infinity $z=1$ (or $r\rightarrow \infty$ in the radial coordinates) can be cast as
\bea
u_{11}(z) &=& Hl(1-t,\alpha\beta - q;\alpha,\beta,\delta,\gamma;1-z)\label{eqn:InfMP71}\,,\\
u_{12}(z) &=& (1-z)^{1-\delta} \times\\ 
&&Hl(1-t,((1-t)\gamma + \epsilon)(1-\delta)+\alpha\beta-q;\alpha+1-\delta,\beta+1-\delta,2-\delta,\gamma;1-z)\label{eqn:InfMP72}\,,\nonumber
\eea

with the most general solution expressed as a linear combination of the two independent solutions
\bea
\label{eqn:InfMP73}
R(z)= z^{\rho}(z-1)^{\sigma}(z-t)^{\tau}\Big[B_3 u_{11}(z)+ B_4 u_{12}(z)\Big]\,.
\eea
with constants $B_3, B_4$.

The differential equation of interest, \eqref{eqn:HeunCan1}, is second-order, so any local solution basis at one singular point can be expressed as a linear combination of the basis at another singular point. We obtain it using the connection formulas of ratios of Wronskians defined on Appendix~\ref{sec:AppeWronsk}. In particular, to expand the local solution at the horizon $z=0$ (e.g. eq.~\eqref{eqn:HorMP73}) around infinity $z=1$ we use the connection formulas $\{C_{11}, C_{12}, C_{21}, C_{22}\}$ defined in eq.~\eqref{eqn:ConFor01}. As a result, we can express the solution at the horizon eq.~\eqref{eqn:HorMP73} as a combination of the local solutions defined at infinity (e.g. eq.~\eqref{eqn:InfMP71} and e.g. eq.~\eqref{eqn:InfMP72})
\bea
\label{eqn:HorInf1}
R(z)=z^{\rho}(z-1)^{\sigma}(z-t)^{\tau}\Big[(B_1 C_{11} + B_2 C_{21}) \,u_{11}(z) + (B_1 C_{12} + B_2 C_{22})\,u_{12}(z)\Big]\,.
\eea
In a similar fashion, we can use the connection formulas $\{D_{11}, D_{12}, D_{21}, D_{22}\}$ defined in eq.~\eqref{eqn:ConFor10} to express the local solution at infinity eq.~\eqref{eqn:InfMP73} as a linear combination of the solutions around the horizon (eg. eq.~\eqref{eqn:HorMP71} and eq.~\eqref{eqn:HorMP72}),
\bea
\label{eqn:InfHor2}
R(z)=z^{\rho}(z-1)^{\sigma}(z-t)^{\tau}\Big[(B_3 D_{11} + B_4 D_{21})u_{01}(z) + (B_3 D_{12} + B_4 D_{22})u_{02}(z)\Big]\,.
\eea
The functions 
\eqref{eqn:HorInf1} and \eqref{eqn:InfHor2} represent the solutions we aimed to derive for the scalar perturbations, constituting our main result. Of course, physically meaningful solutions must satisfy appropriate boundary conditions, which depend on the specific problem at hand — a task we address next as an application. For other problems with different boundary conditions, the most general solutions remain 
\eqref{eqn:HorMP73} and \eqref{eqn:InfMP73}.

\section{Love Numbers for Myers Perry D=7}
We now apply the results obtained of section~\ref{sec:3} to compute the scalar tidal deformation coefficient for the seven-dimensional Myers-Perry black hole. The full spectrum is obtained, including logarithmic running for specific cases. For a detailed discussion on scalar tidal deformation for rotating higher-dimensional black hole solutions, see Refs. ~\cite{Kol:2011vg, Hui:2020xxx, Hui:2021vcv, Charalambous:2021kcz, Charalambous:2021mea, Hui:2022vbh, Charalambous:2022rre, Charalambous:2023jgq, Perry:2023wmm, Perry:2024vwz, Glazer:2024eyi}. 
\subsection{Definition}
For the sake of clarity, let us define the Love numbers relevant to our analysis. Consider $\hat{\ell}=\ell/(D-3)$, where $D$ refers to the dimensionality of spacetime. 
We denote by \( \lambda_{\hat{\ell}, m} \) the tidal response coefficient that characterizes 
the deformation induced in an astrophysical object, such as a black hole, by an external field. 
The complex coefficient \( \lambda_{\hat{\ell}, m} \) encodes both conservative (real) and dissipative (imaginary) components, 
corresponding respectively to the non-dissipative and absorptive parts of the response. The calculation follows the procedure outlined in Refs.~\cite{Kol:2011vg,Hui:2020xxx,Hui:2021vcv,Charalambous:2021kcz,Charalambous:2021mea,Hui:2022vbh,Charalambous:2022rre,Charalambous:2023jgq,Perry:2023wmm,Perry:2024vwz,Glazer:2024eyi}. 
The starting point is the Klein--Gordon equation~\eqref{eqn:KG1}. 
In cases with enhanced symmetry, one can employ a suitable ansatz to obtain a separable equation. 
A notable example is given by \eqref{eqn:MPansatz}.
The radial function satisfies  in-boundary condition on the horizon 
\bea
\label{eqn:BCHor}
R(r)=cons \times (r-r_{+})^{-i{\alpha}_{+}}\,, & {\alpha}_{+}>0\,, & r\rightarrow r_{+}\,,
\eea
where $\alpha_+$ is the characteristic (monodrmoy)parameter.
 At asymptotic infinity, including higher dimensional spacetimes
\bea
\label{eqn:BCInf}
R(r)/r^{\ell} \rightarrow 1\,, \qquad \text{for} \qquad r\rightarrow \infty\,.
\eea
By analyzing the asymptotic expansion of the solution, we obtain the tidal Love number response coefficient that characterizes the object's tidal deformation
\bea
\label{eqn:TidalResp}
R(r)\rightarrow R_{\infty} \,(r^{\ell} + {\lambda}_{\hat{\ell}, m} \, \, r^{-\ell - 2N})\,,
\eea
given by ${\lambda}_{\hat{\ell}, m}$. From our definition, in general, ${\lambda}_{\hat{\ell}, m} \in \mathbb{C}$ -- the imaginary part arises from horizon dissipation, and the real part (which vanishes for 4D Schwarzschild) corresponds to static tidal rigidity.

\subsection{Definition}

We are now in a position to impose the appropriate boundary conditions to extract the Love number from the derived solutions.  
Regularity on the event horizon, at \( z = 0 \) in Eq.~  \eqref{eqn:BCHor}, implies fixing $B_2=0$ such that
\bea
\label{eqn:HeunHor}
R(z)=z^{\rho}(z-1)^{\sigma}(z-t)^{\tau}B_1 u_{01}(z)
\eea

 The solution in the proximity at the boundary \( z = 1 \), preserving the boundary conditions of Eq.~\eqref{eqn:BCInf}, can written as a linear combination 
\begin{align}
\label{eqn:HeunInf}
R(z)=z^{\rho}(z-1)^{\sigma}(z-t)^{\tau}\Big[B_1 (C_{11} \, u_{11}+C_{12}\,u_{12})\Big]\,,
\end{align}
where the $\{C_{11},C_{12}\}$ are familiar connection formulas defined in eq.~\eqref{eqn:ConFor01}. The latter is the expression we are going to employ for the scalar tidal response coefficient.  We have taken our horizon eigenfunction eq.~\eqref{eqn:HeunHor} and express it as a linear combination of our eigenfunctions at infinity, e.g. eq.~\eqref{eqn:HeunInf} by using the identity \eqref{eqn:HeunCon01}. That's why the overall coefficient $B_1$ is kept.
The expansion of our solution (e.g.~\eqref{eqn:HeunInf}) at $r\rightarrow \infty$, due to the presence of both the field redefinition and Wronskians connection terms, makes complicated a clear separation of both the $r^{\hat{\ell}}$ and $r^{-\hat{\ell}-2N}$ branches. Due to this issue, we chose to derive the Love numbers in the small $a << 1$ limit. However, we expect the use of different mathematical techniques and properties would lift this obstruction and allow us to get a Love number for any value of $a$.

\subsection{Tidal Love Numbers}
In this subsection, we present the derivation of the tidal Love numbers in the small-spin limit $a \ll 1$,.
First, we consider the approximate values of the Myers-Perry $D=7$ black hole horizons
\bea
\label{eqn:HorExpan}
&&r_{+} \approx r_{s}\Big(1 - \frac{a^2}{4 r_{s}^{2}}-\frac{7 a^4}{32 r^{4}_{s}}-\frac{39a^{6}}{128r^{6}_{s}}+\mathcal{O}\Big(\frac{a^8}{r^{8}_{s}}\Big)\Big)\,,\\
&&r_{-} \approx a\Big(1 + \frac{a^4}{2 r^{4}_{s}} + \mathcal{O}\Big(\frac{a^8}{r^{8}_{s}}\Big)\Big)\,,\\
&&r_{3} \approx i r_{s}\Big(1 + \frac{a^2}{4 r^{2}_{s}} - \frac{7a^4}{32 r^{4}_{s}} + \frac{39 a^6}{128 r^{6}_{s}}+\mathcal{O}\Big(\frac{a^8}{r^{8}_{s}}\Big)\Big)\,.
\eea

Second, we perform an expansion of the local Heun solution \eqref{eqn:HeunInf} in the limit $r \to \infty$ and for small spin $a \ll 1$. And finally, considering $\eqref{eqn:Mobius2}$, the solution \eqref{eqn:HeunInf} in  the small spin regime $a\ll 1$ yields
\bea
R(z(r))=R_{11} + R_{12}
\label{eqn:HorInf2}\,,
\eea
where
\bea
R_{11} = B_{1}C_{11}z^{\rho}(z-1)^{\sigma}(z-t)^{\tau}u_{11}(z)\label{eqn:HorInf211}\,,\\
\approx B_1 \Bigg(r^{\ell}(N_{1}+\mathcal{O}(r^{-2})) + r^{\ell}\Big(\frac{N_2}{r^4}+\mathcal{O}(r^{-8})\Big)\times \nonumber\\
\hypergeom{2}{1}\left[\begin{array}{c}
1- \hat{\ell},~~1 - \hat{\ell}\\
2
\end{array}\Big\rvert \,1-\frac{r_{s}^{4}}{r^4}\,\right] \Bigg)\label{eqn:HorInfExpan211}\,,
\eea
and
\bea
R_{12} = B_{1}C_{12}z^{\rho}(z-1)^{\sigma}(z-t)^{\tau}u_{12}(z)\label{eqn:HorInf22}\,,\\
\approx  B_1 \Bigg(r^{\ell}\Big(\frac{N_3}{r^4}+\mathcal{O}(r^{-6})\Big)\hypergeom{2}{1}\left[\begin{array}{c}
1- \hat{\ell},~~1 - \hat{\ell}\\
2
\end{array}\Big\rvert \,1-\frac{r_{s}^{4}}{r^{4}}\,\right]+\nonumber\\
r^{\ell}\Big(\frac{N_4}{r^2}+\mathcal{O}(r^{-4})\Big)\hypergeom{2}{1}\left[\begin{array}{c}
-\hat{\ell},~~-\hat{\ell}\\
1
\end{array}\Big\rvert \,1-\frac{r_{s}^{4}}{r^{4}}\,\right] \Bigg)\label{eqn:HorInfExpan221}\,,
\eea
where $\{N_1, N_2, N_3, N_4\}$ are constants expressed in terms of the black hole mass and angular momentum parameters $\{r_s, a\}$. Comparing the expansion \eqref{eqn:HorInf2}, with $\hat{\ell} = \ell/4$, we can identify the growing and decaying contributions, allowing us to extract the tidal response coefficient as defined in \eqref{eqn:TidalResp}.

It is important to consider different cases depending on the specific values of $\hat{\ell}$.

\subsubsection{$\hat{\ell}$ neither integer nor half-integer}
When $\hat{\ell}$ is neither integer nor half-integer ($\hat{\ell}=1/4,3/4,\cdots$), our tidal response coefficient takes the form
\begin{tcolorbox}[colframe=white,arc=0pt,colback=greyish2]
\bea
{\lambda}_{\hat{\ell},m}= \frac{\Gamma(-2\hat{\ell}-1)\Gamma(\hat{\ell}+1)^{2}}{\Gamma(-\hat{\ell})^{2}\Gamma(2\hat{\ell}+1)}\Bigg[1-\frac{3 a^2}{2r^{2}_{s}} - E \frac{a^5}{2 r_s^5}\Bigg]+\mathcal{O}(a^6)\,, && \hat{\ell}=\frac{\ell}{4}\,,
\label{eqn:LoveReals}
\eea
\end{tcolorbox}
where
\bea
E=\frac{-288 + 24i (12\hat{\ell} - 5)\pi - i m^2 \pi^3}{128}\,.
\eea
We emphasize the fact that this expression is valid only for odd values of $\ell$. Equally important, it reduces to the Schwarzschild- Tangherlini rational value case of $\hat{\ell}$ obtained at \cite{Hui:2020xxx}.
\subsubsection{$\hat{\ell}$ integer or half-integer}
For $\hat{\ell}$ integer and half-integer values, the hypergeometric function factors in eqs.~\eqref{eqn:HorInfExpan211} and \eqref{eqn:HorInfExpan221} show special properties. Now we consider the cases where $c-a-b$ are integers. In that context, we should recall the next identity \cite{Bateman:100233} considering $c=a+b+l$, where $l \in {\mathbb{Z}}_+^0$,
\bea
\label{eqn:HeunAnalCon}
\hypergeom{2}{1}\left[\begin{array}{c}
a,~~b\\
a + b + l
\end{array}\Big\rvert \,x\,\right]=\frac{\Gamma(l)\Gamma(a+b+l)}{\Gamma(a+l)\Gamma(b+l)}\sum^{l-1}_{n=0} \frac{(a)_{n}(b)_{n}}{(1-l)_{n}n!}\Big(1-x\Big)^{n}\nonumber\\
+\Big(1-x\Big)^{l}(-1)^{l}\frac{\Gamma(a+b+l)}{\Gamma(a)\Gamma(b)}\sum^{\infty}_{n=0} \frac{(a+l)_{n}(b+l)_{n}}{n!(n+l)!}\Big({\kappa}_{n}-\log\Big(1-x\Big)\Big)(1-x)^{n}\,,
\eea
with
\bea
{\kappa}_{n}=\psi(n+1)+\psi(n+1+l)-\psi(a+n+l)-\psi(b+n+l)\,.
\eea
and $\psi(a)$ the digamma function eq.~\eqref{eqn:digamma}.
Then we apply \eqref{eqn:HeunAnalCon} to the function 
\bea
\hypergeom{2}{1}\left[\begin{array}{c}
1- \hat{\ell},~~1 - \hat{\ell}\\
2
\end{array}\Big\rvert \,1-\frac{r_{s}^{4}}{r^4}\,\right]
\eea
where
\bea
l=2\hat{\ell}\,, a=1-\hat{\ell}\,, b=1-\hat{\ell}\,, c=2\,, x=1-\frac{r_s^4}{r^4}\,,
\eea
given that $c-a-b=2\hat{\ell}$, which applies for both integers and half-integer values of $\hat{\ell}$. We obtain that
\begin{align}
\hypergeom{2}{1}\left[\begin{array}{c}
1- \hat{\ell},~~1 - \hat{\ell}\\
2
\end{array}\Big\rvert \,1-\frac{r_{s}^{4}}{r^4}\,\right]=\frac{\Gamma(2\hat{\ell})\Gamma(2)}{\Gamma(1+\hat{\ell})\Gamma(1+\hat{\ell})}\sum^{2\hat{\ell}-1}_{n=0} \frac{(1-\hat{\ell})_{n}(1-\hat{\ell})_{n}}{(1-2\hat{\ell})_{n}n!}\Big(\frac{r_{s}^{4}}{r^{4}}\Big)^{n}\nonumber\\
+\Big(\frac{r_{s}^{4}}{r^{4}}\Big)^{2\hat{\ell}}(-1)^{2\hat{\ell}}\frac{\Gamma(2)}{\Gamma(1-\hat{\ell})\Gamma(1-\hat{\ell})}\sum^{\infty}_{n=0} \frac{(1+\hat{\ell})_{n}(1+\hat{\ell})_{n}}{n!(n+2\hat{\ell})!}\Big({\kappa}_{n}-\log\Big(\frac{r_{s}^{4}}{r^{4}}\Big)\Big)(\frac{r_{s}^{4}}{r^{4}})^{n}\,,\\
{\kappa}_{n}=\psi(n+1)+\psi(n+1+2\hat{\ell})-\psi(1+\hat{\ell}+n)-\psi(1+\hat{\ell}+n)\,.
\end{align}
Similarly for
\bea
\hypergeom{2}{1}\left[\begin{array}{c}
-\hat{\ell},~~-\hat{\ell}\\
1
\end{array}\Big\rvert \,1-\frac{r_{s}^{4}}{r^{4}}\,\right]\,,
\eea
where
\bea
l=1+2\hat{\ell}\,, a=-\hat{\ell}\,, b=-\hat{\ell}\,, c=1\,, x=1-\frac{r_s^4}{r^4}
\eea
works so that $c-a-b=1+2\hat{\ell}$, again being $\hat{\ell}$ integer or half-integer. 
We get for the second function
\begin{align}
\hypergeom{2}{1}\left[\begin{array}{c}
- \hat{\ell},~~- \hat{\ell}\\
1
\end{array}\Big\rvert \,1-\frac{r_{s}^{4}}{r^4}\,\right]=\frac{\Gamma(1+2\hat{\ell})\Gamma(1)}{\Gamma(1+\hat{\ell})\Gamma(1+\hat{\ell})}\sum^{2\hat{\ell}}_{n=0} \frac{(-\hat{\ell})_{n}(-\hat{\ell})_{n}}{(-2\hat{\ell})_{n}n!}\Big(\frac{r_{s}^{4}}{r^{4}}\Big)^{n}\nonumber\\
+\Big(\frac{r_{s}^{4}}{r^{4}}\Big)^{1+2\hat{\ell}}(-1)^{1+2\hat{\ell}}\frac{\Gamma(1)}{\Gamma(-\hat{\ell})\Gamma(-\hat{\ell})}\sum^{\infty}_{n=0} \frac{(1+\hat{\ell})_{n}(1+\hat{\ell})_{n}}{n!(n+1+2\hat{\ell})!}\Big({\kappa}_{n}-\log\Big(\frac{r_{s}^{4}}{r^{4}}\Big)\Big)(\frac{r_{s}^{4}}{r^{4}})^{n}\,,\\
{\kappa}_{n}=\psi(n+1)+\psi(n+2+2\hat{\ell})-\psi(1+\hat{\ell}+n)-\psi(1+\hat{\ell}+n)\,.
\end{align}
By considering the previous analytic continuation, as well as shifting from $\ell \in \mathbb{R}$ to the $\ell \in \mathbb{N}$ case we recover
\begin{align}
\label{eqn:LoveMP7IntHI}
\lambda_{\hat{\ell},m}=2i(-1+e^{4i\hat{\ell}\pi})\frac{{\hat{\ell}}^{2}\Gamma(-2\hat{\ell})\Gamma(1+\hat{\ell})^{2}}{\pi(1+2\hat{\ell})\Gamma(1-\hat{\ell})^{2}\Gamma(1+2\hat{\ell})}\Big(1-\frac{3a^2}{2 r_s^2}+E \frac{a^5}{2 r_s^5} + \mathcal{O}(a^6)\Big)\log\Big(\frac{r_s}{r}\Big)\,,
\end{align}
where
\bea
E=\frac{-288 + 24i (12\hat{\ell} - 5)\pi - i m^2 \pi^3}{128}\,.
\eea
The above expression can be rewritten in the form
\begin{tcolorbox}[colframe=white,arc=0pt,colback=greyish2]
\bea
\lambda_{\hat{\ell},m} = \frac{(-1)^{2\hat{\ell}}(D-3)\Gamma(\hat{\ell}+1)^{2}}{(2\hat{\ell})!(2\hat{\ell}+1)!\Gamma(-\hat{\ell})^{2}}\Big(1-\frac{3a^2}{2 r_s^2}+E \frac{a^5}{2 r_s^5} + \mathcal{O}(a^6)\Big)\log\Big(\frac{r_s}{r}\Big)\,,
\label{eqn:LoveMP7IntHI2}
\eea
\end{tcolorbox}
using the gamma identities \eqref{eqn:GammaIdent} and the condition.
\bea
\label{eqn:IndetProp}
\lim_{\hat{\ell} \rightarrow \mathbb{Z}_0^+/2} 2i \pi^{-1}\Gamma(-2\hat{\ell})(e^{4\pi i \hat{\ell}} - 1) = \frac{(-1)^{2\hat{\ell}}(D-3)}{2\hat{\ell}!}\,.
\eea
Finally, we find that for $\hat{\ell}$ integer
\begin{tcolorbox}[colframe=white,arc=0pt,colback=greyish2]
\bea
\label{eqn:LoveMP7int}
{\lambda}_{\hat{\ell},m} = 0\,, && \hfill \hat{\ell} \in \mathbb{Z}_{0}^{+}
\eea
\end{tcolorbox}
Note that this reproduces a vanishing Love number, as observed in the case of the five-dimensional Myers–Perry black hole \cite{Rodriguez:2023xjd, Glazer:2024eyi}.

Otherwise, for half-integer $\hat{\ell}$ values, \eqref{eqn:LoveMP7IntHI} does not vanish \cite{Rodriguez:2023xjd, Glazer:2024eyi}. 
As a consistency check, on the spin-less limit ($a=0$), \eqref{eqn:LoveMP7IntHI2} reduces to the Schwarzschild-Tangherlini case
\begin{align}
\label{eqn:LoveSchHI}
\lambda_{\hat{\ell}}=\frac{(-1)^{2\hat{\ell}}(D-3)\Gamma(\hat{\ell}+1)^{2}}{(2\hat{\ell})!(2\hat{\ell}+1)!\Gamma(-\hat{\ell})^{2}}\log\Big(\frac{r_s}{r}\Big)\,.
\end{align}
This expression yields a vanishing Love number for $\hat{\ell} \in \mathbb{Z}$, and reproduces the results of \cite{Hui:2020xxx} for the half-integer case.
As shown in \cite{Hui:2020xxx, Rodriguez:2023xjd, Charalambous:2023jgq, Glazer:2024eyi}, the Love numbers exhibit logarithmic running, providing an explicit manifestation of renormalization group flow in classical gravity. In the half-integer case, the Love number depends logarithmically on $r$, reflecting its scale dependence. The relevant scale is set by the Schwarzschild radius $r_s$, while the Boyer–Lindquist coordinate $r$ measures the distance from the horizon at which the tidal response coefficient is evaluated. By fixing the scale $r_s$ now our coefficient expressed in eq.~\eqref{eqn:LoveMP7IntHI2} runs unambiguously.

\subsection{Connection formula: Wronskian and AGT Approach}

We present a brief overview of an alternative approach to deriving the connection formulas, based on the AGT correspondence. As discussed in Sec.~\ref{Sec:Intro}, a new method for computing the connection formulas for the Heun equation and its confluent limits has been recently developed. This relies on key mathematical identities and insights from the AGT correspondence proposed in \cite{Alday:2009aq}. This framework was further developed by Bonelli, Iossa, Panea-Lichtig, and Tanzini in a series of works \cite{Bonelli:2021uvf, Bonelli:2022ten}. The technical details of these derivations are extensive, and we refer the reader to the original literature for a comprehensive treatment.

Here, we focus on presenting the connection formula as a perturbative series expansion within the AGT framework, expressed order by order using the semiclassical conformal blocks $\mathcal{W}(t)$. For convenience, we adopt the notation introduced by Gutperle and Yeo \cite{Gutperle:2025bzr}.



For simplicity, we begin with the Heun equation in its canonical form, Eq.~\eqref{eqn:HeunCan1}. This equation can be equivalently rewritten in its normal form
\begin{align}
\label{eqn:HeunNor}
\frac{d^2 W}{dz^2}+\Big(\frac{{\delta}_0}{z^2}+\frac{{\delta}_1}{(z-1)^2}+\frac{{\delta}_t}{(z-t)^2}+\frac{{\delta}_{\infty} - {\delta}_0 - {\delta}_1 - {\delta}_t}{z(z-1)}+\frac{(t-1)\mathcal{E}}{z(z-1)(z-t)}\Big)W=0\,,
\end{align}
through the introduction of a field redefinition
\bea
\label{eqn:HNorRedef}
w(z)=z^{-\gamma/2}(z-1)^{-\delta/2}(z-t)^{-\epsilon/2}W(z)\,.
\eea
The relation between the characteristic exponent parameters of \eqref{eqn:HeunCan1} and \eqref{eqn:HeunNor} is given by
\bea
\label{eqn:CanToNor}
\delta_0 = \frac{\gamma}{4}(\gamma-2)\,, && \delta_1 = \frac{\delta}{4}(\delta - 2)\,,\\
\delta_t = \frac{\epsilon}{4}(\epsilon - 2)\,, && \delta_{\infty} = -\alpha\beta + \frac{1}{4}(-2+\gamma + \delta + \epsilon)(\gamma + \delta + \epsilon)\,,
\eea
and the accessory parameter defined as
\begin{align}
\label{eqn:AccsParam}
\mathcal{E}=\frac{2q-2t\alpha\beta+\gamma\epsilon(t-1)+t \delta\epsilon}{2(t-1)}\,.
\end{align}
The proposed connection formula, which links the Frobenius solutions at $z=0$ and $z=1$, can then be written as
\begin{align}
\label{eqn:HeunCon}
w_{j}^{[0]}(z)=\sum_{j'}C(j{\theta}_{0},j'{\theta}_{1},\sigma)w_{j'}^{[1]}(z)\,, && j,j'=\pm\,,
\end{align}
\begin{align}
C(j{\theta}_{0},j'{\theta}_{1},\sigma)=\frac{\Gamma(1-2j{\theta}_{0})\Gamma(2j'{\theta}_{1})}{\Gamma(\frac{1}{2}-2j{\theta}_{0}+2j'{\theta}_{1}\pm \sigma)}\exp{\Big[\frac{1}{2}(j'{\partial}_{{\theta}_1}-j{\partial}_{{\theta}_0})\mathcal{W}(t)\Big]}\,,
\end{align}
given that $\Gamma(a\pm b)=\Gamma(a+b)\Gamma(a-b)$. 
Here the monodromy parameters $\theta_i$ associated for each regular singular point of the equation are defined in terms of eq.~\eqref{eqn:HeunNor} characteristic exponent parameters given the following relationship
\begin{align}
\label{eqn:ExpToMonodromy}
    {\delta}_{i}=\frac{1}{4}-{\theta}_{i}^{2}\,,
\end{align}
where $i$ is the label referring the fourth singular points $\{0,1,t,\infty\}$.
Both $\sigma$ and $\omega$ here are set by relations involving the series expansion of semiclassical conformal block having as center the value of the singular point $t$, considering this to be small but always $|t| \geq  1$.

The so-called semiclassical conformal block $\mathcal{W}(t)$ is usually expressed as perturbative expansion respect to parameter $t$,
\bea
    \label{eqn:SCB}
    \mathcal{W}(t)&=&({\delta}_{\infty}-{\delta}_{\sigma}-{\delta}_{t})\ln{t}+\sum_{k=1}^{\infty}\mathcal{W}_{k}t^{-k}\,,\\
    \mathcal{W}_{1}&=&\frac{({\delta}_{\sigma}-{\delta}_{0}+{\delta}_{1})({\delta}_{\sigma}-{\delta}_{\infty}+{\delta}_{t})}{2{\delta}_{\sigma}}\,,\\
    \mathcal{W}_{2}&=&\frac{({\delta}_{\sigma}-{\delta}_{0}+{\delta}_{1})^{2}({\delta}_{\sigma}-{\delta}_{\infty}+{\delta}_{t})^{2}}{8{\delta}_{\sigma}^{2}}\Big(\frac{1}{{\delta}_{\sigma}-{\delta}_{0}+{\delta}_{1}}+\frac{1}{{\delta}_{\sigma}-{\delta}_{\infty}+{\delta}_{t}}-\frac{1}{2{\delta}_{\sigma}}\Big)\nonumber\\
    &&+\frac{\Big({\delta}_{\sigma}^{2}+2{\delta}_{\sigma}({\delta}_{0}+{\delta}_{1})-3({\delta}_{0}-{\delta}_{1})^{2}\Big)\Big({\delta}_{\sigma}^{2}+2{\delta}_{\sigma}({\delta}_{\infty}+{\delta}_{t})-3({\delta}_{\infty}-{\delta}_{t})^{2}\Big)}{16{\delta}_{\sigma}^{2}(4{\delta}_{\sigma}+3)}\,.
\eea 
The AGT approach and the Wronskian method \cite{Hatsuda:2021gtn, Noda:2022zgk}, are connected in a subtle form. On \cite{Lisovyy:2022flm}, Lisovvy and Naidiuk show how the connection coefficient expressed as a ratio of Wronskians can be presented as a perturbative expansion. Based on the works of Darboux \cite{darboux1878memoire} and Schafke and Schmidt \cite{schafke1980connection}, the equivalence between both methods is proven for both the Heun and Confluent Heun equation. For additional evidence on the application of the AGT approach, refer to \cite{Bonelli:2021uvf} \cite{Consoli:2022eey} \cite{Jia:2024zes} \cite{Gutperle:2025bzr}.

\section{Discussion}
\label{sec:Discuss}

In this paper, we investigated black hole perturbations focusing on Heun differential equations, emphasizing Frobenius power-series solutions near regular singularities and their connection formulas. Our primary goal was to demonstrate how to solve a generalized Heun equation \eqref{eqn:gen} leveraging the analytic structure of its singular points, which lie in clines in the complex plane.\\
\indent Our work is motivated by the many problems in physics that involve (generilized) Heun differential equations. Although these equations possess well-known local solutions that can be formally expressed near their regular singular points, they frequently become analytically intractable due to the high number of singularities and the challenges involved in deriving the corresponding connection formulas.\\
\indent The main conclusion of this paper is that, in all cases we studied, the geometric arrangement of an ODE’s singular points along a cline enables us to exploit their invariance under Möbius transformations, thereby reducing the problem to a tractable form. An important step in this approach is reduction of generalized Heun equations to the standard Heun equation. This prompts the question of whether all solutions to black hole perturbation problems can ultimately be expressed in terms of Heun functions, along with the well-established connection formulas for the Heun equation.\\
\indent We considered the following properties of the black hole perturbation equations. First, central to our approach is the concept of a cline in the complex plane, which organizes the singular points of the differential equations and remains invariant under Möbius transformations. Second, building on the cline structure—identified, to the best of our knowledge, for the first time at black hole horizons—we carry out a systematic reduction and relocation of the poles in the differential equations to derive explicit representations of the solutions. Thus, we confirm that the black hole perturbation problems— characterized by the generalized Heun equations \eqref{eqn:gen}— can be solved by exploiting the analytic structure of their singular points (black hole horizons) organized as clines in the complex plane. This is achieved by mapping the original equation into coordinates \( r \rightarrow x \propto r^2 \), making the Möbius covariance explicit. These transformations preserve the Fuchsian character of the equation \cite{ronveaux1995heun, slavjanov2000special}.\\
\indent As a practical exercise, we solved the radial sector of the scalar wave perturbation of the Myers-Perry black hole in seven dimensions \eqref{eqn:KGMPStatic7d2}. Then we derived the eigenfunctions for the resultant Heun equation, expressed as local Heun functions around the event horizon $z=0$ \eqref{eqn:HorMP72} and asymptotic infinity $z=1$ \eqref{eqn:InfMP72}. Thus we obtained an exact solution for our equation.\\
\indent We used the result obtained in eqs.~\eqref{eqn:HorMP73} and \eqref{eqn:InfMP73} to derive the static Love numbers. Under suitable boundary conditions (eqs.~\eqref{eqn:BCHor} and \eqref{eqn:BCInf}) proper eigenfunctions were found around the event horizon (eq.~\eqref{eqn:HeunHor}) and infinity (eq.~\eqref{eqn:HeunInf}). With the aid of the Wronskian connection formulas introduced in Appendix~\ref{sec:AppeWronsk}, the connection between the local solution on the horizon and at infinity is obtained. 
As a test, we derived the Love Numbers for the Myers-Perry black hole in 7 dimensions.\\
\indent Regarding the Love number calculation, we derived the complete spectrum of static tidal response coefficients for the seven-dimensional Myers-Perry black hole. Those include the tidal response response coefficient for $\hat{\ell}$ integer eq.~\eqref{eqn:LoveMP7int}, $\hat{\ell}$ half-integer eq.~\eqref{eqn:LoveMP7IntHI2}, and for the non-integer nor half-integer $\hat{\ell}$ value eq.~\eqref{eqn:LoveReals}. As seen on previous results both for rotating \cite{Charalambous:2023jgq, Rodriguez:2023xjd, Glazer:2024eyi} and non-rotating \cite{Hui:2020xxx} black hole solutions, we recover the logarithmic running behavior for the $\hat{\ell}$ half-integer value. For comparison with the case in five dimensions with angular momentum parameters set equal, we present on Appendix~\ref{sec:MP5Love} its Love number spectrum for the small angular momentum case, based on the calculations presented in \cite{Rodriguez:2023xjd}. Similarly we determined that on the zero-angular momentum limit $a=0$, all our results agree for all $\hat{\ell}$ values with those derived in  \cite{Hui:2020xxx}. \\
\indent The cline structure of the regular singular points can be used to diagnose the reducibility of our original Fuchsian equation under a proper $z \propto r^m$ mapping, dependent on the poles $m$-th multiplicity. In general, the conjecture here presented is susceptible of extension as far as the Fuchsian nature of the system is not altered. This is compatible with cases beyond Myers-Perry, like rotating BTZ \cite{Guica:2014dfa, Dias:2019ery, Kajuri:2020bvi} or Kerr-AdS \cite{Cardoso:2006wa, Ishii:2022lwc, Graf:2025hhd} (see Fig.~\ref{fig:HorBTZKAdS}).
Similar problems in quantum mechanics and other fields of physics shall be able to benefit from it, with the potential obstruction on having deeper problems on finding the appropriate rational map for the case at hand.
\begin{figure}[h!]
\centering
\begin{subfigure}{.45\textwidth}
    \centering
    \includegraphics[width=\linewidth]{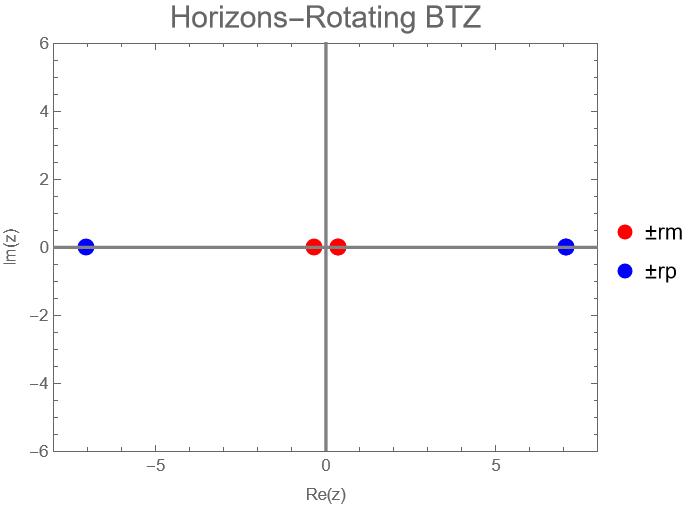}
\end{subfigure}
\hfill
\begin{subfigure}{.45\textwidth}
    \centering
    \includegraphics[width=\linewidth]{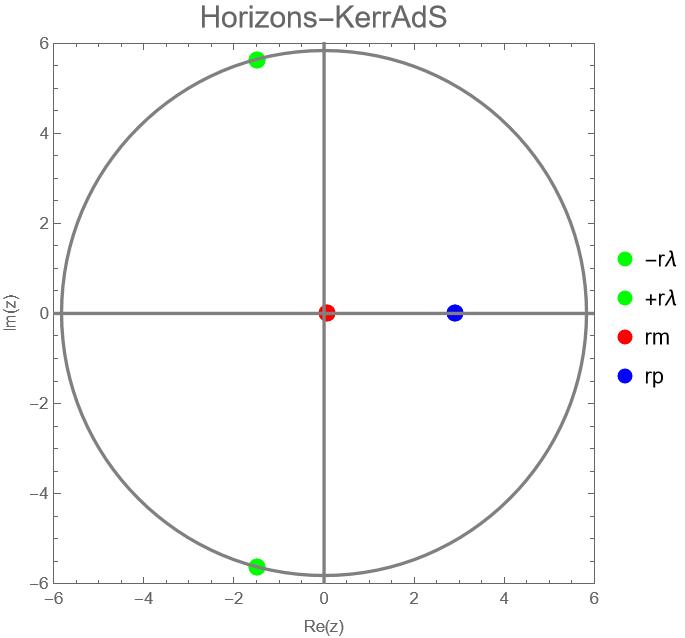}
\end{subfigure}
\caption{Regular singular points for the KG equations of rotating BTZ and Kerr-AdS.}
\label{fig:HorBTZKAdS}
\end{figure}
\\
\indent We now comment on future directions. Our current study has been limited to Heun equations with only regular singular points (Fuchsian systems). However, black hole systems of interest, including Schwarzschild or Kerr, feature irregular singular points at infinity. As a next step in our analysis, we aim to generalize our study to the confluent limits of the cases considered. This includes the time-dependent scalar equations ($\omega\ne 0$) and their extension to full gravitational perturbations. Similarly, obtaining exact solutions for other black-brane species and black string cases would represent a natural next step. At a fundamental level, it would be valuable to test the universality of our proposal against alternative approaches, such as those in \cite{Bonelli:2021uvf, Jia:2024zes}. An interesting direction for black hole perturbation problems  is to perform comparisons with so called-MST method \cite{Sasaki:2003xr}. The qualitative conclusions drawn from this analysis, even when obtained in specific regimes (such as static cases), may have broader applicability and help clarify the invariance between solutions characterized by the generalized momentum parameter $\nu$ and $-\nu-1$ .

\section*{Acknowledgments}

We would like to thank Adam R. Solomon and Alba Grassi for useful comments on the present work and related topics. We thank also the Mitchell Family Foundation for hosting us during the Cook's Branch workshop where some of the research was carried out. 
MJR is partially supported through the NSF grant PHY-2309270.  LFT's work is supported by the USU Presidential Doctoral Research Fellowship, Howard L. Blood Fellowship and Farrell and Ann Edwards Scholarship.

\appendix

\section{Connection Formulas via Wronskians}
\label{sec:AppeWronsk}
Our overview of the use of Wronskian ratios as connection formulas is based on the theory presented by Olver \cite{olver1997asymptotics} and Ince \cite{ince2012ordinary}, together with their applications in gravitational theory, as in the works of Hatsuda \cite{Hatsuda:2021gtn}, Motohashi \cite{Motohashi:2021zyv} and Noda \cite{Noda:2022zgk}. 

We start with the \textit{generalized Heun} equation and under the action of \eqref{eqn:RatMap}, it is reduced to a \textit{Heun} equation, expressed in its canonical form as
\bea
\label{eqn:HeunEq}
\frac{d^2 y}{dz^2} + \Big[\frac{\gamma}{z} + \frac{\delta}{z-1} + \frac{\epsilon}{z-a}\Big]\frac{dy}{dz} + \frac{\alpha\beta z -q}{z(z-1)(z-a)}y = 0\,.
\eea
We then determine the local Heun functions $Hl$, whose radius of convergence is given by $|z|<\min(1,|a|)$. In particular, we focus on the solutions expanded around $z=0$,
\bea
\label{eqn:HeunZero}
&&y_{01}(z) = Hl(a,q;\alpha,\beta,\gamma,\delta;z)\,,\\
&&y_{02}(z) = z^{1-\gamma}Hl(a,(a\delta + \epsilon)(1-\gamma)+q;\alpha+1-\gamma,\beta + 1 -\gamma,2-\gamma,\delta;z)\,,
\eea
and for $z=1$,
\bea
\label{eqn:HeunOne}
&&y_{11}(z) = Hl(1-a,\alpha\beta - q;\alpha,\beta,\delta,\gamma;1-z)\,,\\
&&y_{12}(z) = Hl(1-a,((1-a)\gamma + \epsilon)(1-\delta)+\nonumber\\
&&\alpha\beta-q;\alpha+1-\delta,\beta+1-\delta,2-\delta,\gamma;1-z)\,.
\eea
The connection formula for solutions around $z=0$ to $z=1$ are given by
\bea
y_{01}(z) = C_{11} \,y_{11}(z) + C_{12}\, y_{12}(z)\label{eqn:HeunCon01}\,,\\
y_{02}(z) = C_{21}\, y_{11}(z) + C_{22}\, y_{12}(z)\label{eqn:HeunCon02}\,,
\eea
with coefficients defined by the ratio of the Wronskians $W_{z}[f,g]=f g'- g f'$,
\bea
\label{eqn:ConFor01}
C_{11}=\frac{W_{z}[y_{01},y_{12}]}{W_{z}[y_{11},y_{12}]}\,,  C_{12}=\frac{W_{z}[y_{01},y_{11}]}{W_{z}[y_{12},y_{11}]}\,,  C_{21}=\frac{W_{z}[y_{02},y_{12}]}{W_{z}[y_{11},y_{12}]}\,, C_{22}=\frac{W_{z}[y_{02},y_{11}]}{W_{z}[y_{12},y_{11}]}\,.
\eea
Similarly we can perform a connection for local functions at $z=1$ towards $z=0$,
\bea
\label{eqn:HeunCon10}
y_{11}(z)=D_{11}y_{01}(z)+D_{12}y_{02}(z)\,,\\
y_{12}(z)=D_{21}y_{01}(z)+D_{22}y_{02}(z)\,,
\eea
namely
\bea
\label{eqn:ConFor10}
D_{11}=\frac{W_{z}[y_{11},y_{02}]}{W_{z}[y_{01},y_{02}]}\,, D_{12}=\frac{W_{z}[y_{11},y_{01}]}{W_{z}[y_{02},y_{01}]}\,, D_{21}=\frac{W_{z}[y_{12},y_{02}]}{W_{z}[y_{01},y_{02}]}\,, D_{22}=\frac{W_{z}[y_{12},y_{01}]}{W_{z}[y_{02},y_{01}]}\,.
\eea
These formulas yield a continuous and analytic solution within the domain of interest, namely for $z \in [0; 1]$.

\section{Love Numbers for \texorpdfstring{$5D$}{5D} Black Holes}
\label{sec:MP5Love}

For completeness, in this appendix we summarize the results for the static Love numbers of five-dimensional black holes following the notation of Ref.~\cite{Hui:2020xxx, Rodriguez:2023xjd}).

\subsection{\texorpdfstring{$5D$}{5D} Myers-Perry Black Hole with  Unequal Spins}
The static Love numbers for the Myers-Perry black hole with all distinct spin parameters, $a\neq b$ are:
\begin{align}
\label{eqn:MP5Love}
{\lambda}_{l,{\tilde{m}}_{L},{\tilde{m}}_{R}}=2(-1)^{l}\frac{(-\hat{l}+2i{\tilde{m}}_{L})_{l+1}(-\hat{l}+2i{\tilde{m}}_{R})_{l+1}}{l!(l+1)!}\log\Big(\frac{r_s}{r}\Big)\,,
\end{align}
where
\begin{align}
{\tilde{m}}_{L}=\frac{a-b}{r_{+} + r_{-}}\frac{m_L}{2}\,, && {\tilde{m}}_{R}=\frac{a+b}{r_{+} - r_{-}}\frac{m_R}{2}\,,\\
m_L = \frac{m_{\phi} - m_{\psi}}{2}\,, && m_R = \frac{m_{\phi} + m_{\psi}}{2}\,, 
\end{align}
noting that
\begin{align}
2r_{\pm}^{2} = \mu - a^2 - b^2 \pm \sqrt{(\mu - a^2 - b^2)^2 - 4 a^2 b^2}\,.
\end{align}
\subsection{Schwarzschild Black Hole}
In the limit $a=b=0$, Eq.~\eqref{eqn:MP5Love} simplifies to
\begin{align}
\label{eqn:Sch5dimHI}
{\lambda}_{l}={\lambda}_{Sch}\log\Big(\frac{r_s}{r}\Big)\,,
\end{align}
where
\begin{align}
{\lambda}_{Sch}=2(-1)^{l}\frac{\Gamma(\hat{l} + 1)^{2}}{l!(l+1)!\Gamma(-\hat{l})^{2}}\,.
\end{align}
\subsection{\texorpdfstring{$5D$}{5D} Myers-Perry Black Hole with Equal Spins}
Similarly, we can set the parameters to the equal-spin case $a=b$,
\begin{align}
{\lambda}_{l,{\tilde{m}}_{R}}=2(-1)^{l}\frac{(-\hat{l})_{l+1}(-\hat{l}+2i{\tilde{m}}_{R})_{l+1}}{l!(l+1)!}\log\Big(\frac{r_s}{r}\Big)\,.
\end{align}
In the small spin regime, where $a \ll 1$, one can expand the function as a series in powers of 
$a$
\begin{align}
{\lambda}_{l,{\tilde{m}}_{R}}\approx {\lambda}_{Sch}\Big(1+\frac{2i m_{R}(\psi(\hat{l}+1)-\psi(-\hat{l}))a}{\sqrt{\mu}}+\mathcal{O}(a^2) \Big)\log\Big(\frac{r_s}{r}\Big)
\end{align}
\section{Mathematical identities}
\label{sec:MathIden}

We examine several mathematical identities that are relevant to this work.

\subsection{Hypergeometric function identities}
Hypergeometric function identities. In that context, we should recall the next identity \cite{Bateman:100233, Derezi_ski_2013} considering $c=a+b+l$, where $l \in {\mathbb{Z}}_+^0$,
\bea
\label{eqn:HAnalConAbstract}
\hypergeom{2}{1}\left[\begin{array}{c}
a,~~b\\
a + b + l
\end{array}\Big\rvert \,x\,\right]=\frac{\Gamma(l)\Gamma(a+b+l)}{\Gamma(a+l)\Gamma(b+l)}\sum^{l-1}_{n=0} \frac{(a)_{n}(b)_{n}}{(1-l)_{n}n!}\Big(1-x\Big)^{n}\nonumber\\
+\Big(1-x\Big)^{l}(-1)^{l}\frac{\Gamma(a+b+l)}{\Gamma(a)\Gamma(b)}\sum^{\infty}_{n=0} \frac{(a+l)_{n}(b+l)_{n}}{n!(n+l)!}\Big({\kappa}_{n}-\log\Big(1-x\Big)\Big)(1-x)^{n}\,,
\eea
with
\bea
{\kappa}_{n}=\psi(n+1)+\psi(n+1+l)-\psi(a+n+l)-\psi(b+n+l)\,.
\eea
being $psi(a)$ the digamma function defined as
\bea
\label{eqn:digamma}
\psi(a) = \frac{\Gamma'(a)}{\Gamma(a)}\,, && a \in \mathbb{C}\,.
\eea
\subsection{Gamma function identities}
Useful identities for the Gamma functions can be found at \cite{Bateman:100233}. Consider that $z \notin \mathbb{Z}$.
\bea
\label{eqn:GammaIdent}
&&\Gamma(1+z)=z\Gamma(z)\,,\\
&&\Gamma(1-z) \Gamma(z) = -z \Gamma(-z)\Gamma(z) = \frac{\pi}{\sin{\pi z}}\,,\\
&&\Gamma(z)\Gamma(-z)=-\pi z^{-1} \csc{\pi z}\,.
\eea

\subsection{Heun local functions}
We follow the definition as presented by Ronveaux et.al.~\cite{ronveaux1995heun}. Given the \textit{Heun equation} on its canonical form \eqref{eqn:HeunCan1}, the \textit{Heun local function} around the $z=0$ singular point is expressed as the Frobenius series solution is defined
\bea
\label{eqn:HeunLocal}
Hl[t,q;\alpha,\beta,\gamma,\delta;z]=\sum_{j=0}^{\infty} c_j z^j\,,  (c_0 \neq 0)\,.
\eea
The coefficients of the series are determined via a three-term recurrence relation
\bea
\label{eqn:RecurRel}
-q c_0 + t\gamma c_1 &=& 0\,,\\
P_j c_{j-1} - (Q_j + q)c_j + R_j c_{j+1} &=& 0\,, (r \geq 1)\,
\eea
where
\bea
\label{eqn:RecurCoef}
&&P_j = (j-1+\alpha)(j-1+\beta)\,,\\
&&Q_j = j[(r-1+\gamma)(1+t)+t\delta+\epsilon]\,,\\
&&R_j = (j+1)(j+\gamma)t\,.
\eea
The characteristic exponent parameters for each regular singular point satisfy the Fuchsian relation
\bea
\label{eqn:FuchsHeun}
\alpha + \beta + 1 = \gamma + \delta + \epsilon\,.
\eea
As a convention, we usually set $c_0 = 1$. The convergence of the series is guaranteed if $|t|>1$ holds.

\addcontentsline{toc}{section}{References}
\bibliography{mybib.bib}

@article{Arnaudo:2025kof,
    author = "Arnaudo, Paolo and Grassi, Alba and Hao, Qianyu",
    title = "{On quivers, spectral networks and black holes}",
    eprint = "2502.01526",
    archivePrefix = "arXiv",
    primaryClass = "hep-th",
    reportNumber = "CERN-TH-2025-015",
    month = "2",
    year = "2025"
}

@article{Aminov:2020yma,
    author = "Aminov, Gleb and Grassi, Alba and Hatsuda, Yasuyuki",
    title = "{Black Hole Quasinormal Modes and Seiberg{\textendash}Witten Theory}",
    eprint = "2006.06111",
    archivePrefix = "arXiv",
    primaryClass = "hep-th",
    reportNumber = "RUP-20-18",
    doi = "10.1007/s00023-021-01137-x",
    journal = "Annales Henri Poincare",
    volume = "23",
    number = "6",
    pages = "1951--1977",
    year = "2022"
}

@article{Camilloni:2023wyn,
    author = "Camilloni, Filippo and Harmark, Troels and Orselli, Marta and Rodriguez, Maria J.",
    title = "{Blandford-Znajek jets in MOdified Gravity}",
    eprint = "2307.06878",
    archivePrefix = "arXiv",
    primaryClass = "gr-qc",
    doi = "10.1088/1475-7516/2024/01/047",
    journal = "JCAP",
    volume = "01",
    pages = "047",
    year = "2024"
}

@incollection{Myers:2011yc,
  author       = {Myers, Robert C.},
  editor       = {Horowitz, Gary T.},
  title        = {Myers--Perry black holes},
  booktitle    = {Black Holes in Higher Dimensions},
  publisher    = {Cambridge University Press},
  address      = {Cambridge, UK},
  year         = {2012},
  pages        = {101--133},
  eprint       = {1111.1903},
  archivePrefix= {arXiv},
  primaryClass = {gr-qc}
}

@article{Ishii:2022lwc,
    author = "Ishii, Takaaki and Kaku, Youka and Murata, Keiju",
    title = "{Energy extraction from AdS black holes via superradiance}",
    eprint = "2207.03123",
    archivePrefix = "arXiv",
    primaryClass = "hep-th",
    reportNumber = "RUP-22-14",
    doi = "10.1007/JHEP10(2022)024",
    journal = "JHEP",
    volume = "10",
    pages = "024",
    year = "2022"
}

@article{Graf:2025hhd,
    author = "Graf, Olivier",
    title = "{Stationary gravitational modes on Kerr-anti-de Sitter spacetimes}",
    eprint = "2506.18524",
    archivePrefix = "arXiv",
    primaryClass = "math.AP",
    month = "6",
    year = "2025"
}

@article{Cardoso:2006wa,
    author = "Cardoso, Vitor and Dias, Oscar J. C. and Yoshida, Shijun",
    title = "{Classical instability of Kerr-AdS black holes and the issue of final state}",
    eprint = "hep-th/0607162",
    archivePrefix = "arXiv",
    doi = "10.1103/PhysRevD.74.044008",
    journal = "Phys. Rev. D",
    volume = "74",
    pages = "044008",
    year = "2006"
}

@article{Chen:2017ofv,
    author = "Chen, Baoyi and Stein, Leo C.",
    title = "{Separating metric perturbations in near-horizon extremal Kerr}",
    eprint = "1707.05319",
    archivePrefix = "arXiv",
    primaryClass = "gr-qc",
    doi = "10.1103/PhysRevD.96.064017",
    journal = "Phys. Rev. D",
    volume = "96",
    number = "6",
    pages = "064017",
    year = "2017"
}

@misc{NIST:DLMF,
         key = "{\relax DLMF}",
       title = "{\it NIST Digital Library of Mathematical Functions}",
howpublished = "\url{https://dlmf.nist.gov/}, Release 1.2.4 of 2025-03-15",
         url = "https://dlmf.nist.gov/",
        note = "F.~W.~J. Olver, A.~B. {Olde Daalhuis}, D.~W. Lozier, B.~I. Schneider,
                R.~F. Boisvert, C.~W. Clark, B.~R. Miller, B.~V. Saunders,
                H.~S. Cohl, and M.~A. McClain, eds."}

@article{Guica:2014dfa,
    author = "Guica, Monica and Ross, Simon F.",
    title = "{Behind the geon horizon}",
    eprint = "1412.1084",
    archivePrefix = "arXiv",
    primaryClass = "hep-th",
    doi = "10.1088/0264-9381/32/5/055014",
    journal = "Class. Quant. Grav.",
    volume = "32",
    number = "5",
    pages = "055014",
    year = "2015"
}

@article{Dias:2019ery,
    author = "Dias, Oscar J. C. and Reall, Harvey S. and Santos, Jorge E.",
    title = "{The BTZ black hole violates strong cosmic censorship}",
    eprint = "1906.08265",
    archivePrefix = "arXiv",
    primaryClass = "hep-th",
    doi = "10.1007/JHEP12(2019)097",
    journal = "JHEP",
    volume = "12",
    pages = "097",
    year = "2019"
}

@article{Kajuri:2020bvi,
    author = "Kajuri, Nirmalya",
    title = "{Bulk reconstruction in rotating BTZ black hole}",
    eprint = "2012.07151",
    archivePrefix = "arXiv",
    primaryClass = "hep-th",
    doi = "10.1103/PhysRevD.103.066019",
    journal = "Phys. Rev. D",
    volume = "103",
    number = "6",
    pages = "066019",
    year = "2021"
}

@article{Ivanov:2022qqt,
    author = "Ivanov, Mikhail M. and Zhou, Zihan",
    title = "{Vanishing of Black Hole Tidal Love Numbers from Scattering Amplitudes}",
    eprint = "2209.14324",
    archivePrefix = "arXiv",
    primaryClass = "hep-th",
    doi = "10.1103/PhysRevLett.130.091403",
    journal = "Phys. Rev. Lett.",
    volume = "130",
    number = "9",
    pages = "091403",
    year = "2023"
}

@article{Dias:2009ex,
    author = "Dias, Oscar J. C. and Reall, Harvey S. and Santos, Jorge E.",
    title = "{Kerr-CFT and gravitational perturbations}",
    eprint = "0906.2380",
    archivePrefix = "arXiv",
    primaryClass = "hep-th",
    doi = "10.1088/1126-6708/2009/08/101",
    journal = "JHEP",
    volume = "08",
    pages = "101",
    year = "2009"
}

@article{Bredberg:2009pv,
    author = "Bredberg, Irene and Hartman, Thomas and Song, Wei and Strominger, Andrew",
    title = "{Black Hole Superradiance From Kerr/CFT}",
    eprint = "0907.3477",
    archivePrefix = "arXiv",
    primaryClass = "hep-th",
    doi = "10.1007/JHEP04(2010)019",
    journal = "JHEP",
    volume = "04",
    pages = "019",
    year = "2010"
}

@article{Compere:2017hsi,
    author = "Comp{\`e}re, Geoffrey and Fransen, Kwinten and Hertog, Thomas and Long, Jiang",
    title = "{Gravitational waves from plunges into Gargantua}",
    eprint = "1712.07130",
    archivePrefix = "arXiv",
    primaryClass = "gr-qc",
    doi = "10.1088/1361-6382/aab99e",
    journal = "Class. Quant. Grav.",
    volume = "35",
    number = "10",
    pages = "104002",
    year = "2018"
}

@article{Porfyriadis:2014fja,
    author = "Porfyriadis, Achilleas P. and Strominger, Andrew",
    title = "{Gravity waves from the Kerr/CFT correspondence}",
    eprint = "1401.3746",
    archivePrefix = "arXiv",
    primaryClass = "hep-th",
    doi = "10.1103/PhysRevD.90.044038",
    journal = "Phys. Rev. D",
    volume = "90",
    number = "4",
    pages = "044038",
    year = "2014"
}

@article{Yang:2013uba,
    author = "Yang, Huan and Zimmerman, Aaron and Zengino{\u{g}}lu, An{\i}l and Zhang, Fan and Berti, Emanuele and Chen, Yanbei",
    title = "{Quasinormal modes of nearly extremal Kerr spacetimes: spectrum bifurcation and power-law ringdown}",
    eprint = "1307.8086",
    archivePrefix = "arXiv",
    primaryClass = "gr-qc",
    doi = "10.1103/PhysRevD.88.044047",
    journal = "Phys. Rev. D",
    volume = "88",
    number = "4",
    pages = "044047",
    year = "2013"
}

@article{Castro:2021csm,
    author = "Castro, Alejandra and Godet, Victor and Sim{\'o}n, Joan and Song, Wei and Yu, Boyang",
    title = "{Gravitational perturbations from NHEK to Kerr}",
    eprint = "2102.08060",
    archivePrefix = "arXiv",
    primaryClass = "hep-th",
    doi = "10.1007/JHEP07(2021)218",
    journal = "JHEP",
    volume = "07",
    pages = "218",
    year = "2021"
}

@article{Cvetic:2024dvn,
    author = "Cveti{\v{c}}, Mirjam and Rodr{\'\i}guez, Nelson Hern{\'a}ndez and Rodriguez, Maria J. and Varela, Oscar",
    title = "{Kerr effective black hole geometries in supergravity}",
    eprint = "2406.10458",
    archivePrefix = "arXiv",
    primaryClass = "hep-th",
    doi = "10.1103/zp85-xym1",
    journal = "Phys. Rev. D",
    volume = "112",
    number = "2",
    pages = "026007",
    year = "2025"
}

@article{Bautista:2023sdf,
    author = "Bautista, Yilber Fabian and Bonelli, Giulio and Iossa, Cristoforo and Tanzini, Alessandro and Zhou, Zihan",
    title = "{Black hole perturbation theory meets CFT2: Kerr-Compton amplitudes from Nekrasov-Shatashvili functions}",
    eprint = "2312.05965",
    archivePrefix = "arXiv",
    primaryClass = "hep-th",
    doi = "10.1103/PhysRevD.109.084071",
    journal = "Phys. Rev. D",
    volume = "109",
    number = "8",
    pages = "084071",
    year = "2024"
}

@article{Kunduri:2006qa,
    author = "Kunduri, Hari K. and Lucietti, James and Reall, Harvey S.",
    title = "{Gravitational perturbations of higher dimensional rotating black holes: Tensor perturbations}",
    eprint = "hep-th/0606076",
    archivePrefix = "arXiv",
    reportNumber = "DAMTP-2006-46",
    doi = "10.1103/PhysRevD.74.084021",
    journal = "Phys. Rev. D",
    volume = "74",
    pages = "084021",
    year = "2006"
}

@article{Castro:2013lba,
    author = "Castro, Alejandra and Lapan, Joshua M. and Maloney, Alexander and Rodriguez, Maria J.",
    title = "{Black Hole Scattering from Monodromy}",
    eprint = "1304.3781",
    archivePrefix = "arXiv",
    primaryClass = "hep-th",
    doi = "10.1088/0264-9381/30/16/165005",
    journal = "Class. Quant. Grav.",
    volume = "30",
    pages = "165005",
    year = "2013"
}

@article{Castro:2013kea,
    author = "Castro, Alejandra and Lapan, Joshua M. and Maloney, Alexander and Rodriguez, Maria J.",
    title = "{Black Hole Monodromy and Conformal Field Theory}",
    eprint = "1303.0759",
    archivePrefix = "arXiv",
    primaryClass = "hep-th",
    doi = "10.1103/PhysRevD.88.044003",
    journal = "Phys. Rev. D",
    volume = "88",
    pages = "044003",
    year = "2013"
}

@article{CarneirodaCunha:2015hzd,
    author = "Carneiro da Cunha, Bruno and Novaes, F{\'a}bio",
    title = "{Kerr Scattering Coefficients via Isomonodromy}",
    eprint = "1506.06588",
    archivePrefix = "arXiv",
    primaryClass = "hep-th",
    doi = "10.1007/JHEP11(2015)144",
    journal = "JHEP",
    volume = "11",
    pages = "144",
    year = "2015"
}

@article{Bhattacharyya:2015dva,
    author = "Bhattacharyya, Sayantani and De, Anandita and Minwalla, Shiraz and Mohan, Ravi and Saha, Arunabha",
    title = "{A membrane paradigm at large D}",
    eprint = "1504.06613",
    archivePrefix = "arXiv",
    primaryClass = "hep-th",
    reportNumber = "TIFR-TH-15-11",
    doi = "10.1007/JHEP04(2016)076",
    journal = "JHEP",
    volume = "04",
    pages = "076",
    year = "2016"
}

@book{olver1997asymptotics,
  title={Asymptotics and special functions},
  author={Olver, Frank},
  year={1997},
  publisher={AK Peters/CRC Press}
}

@book{ince2012ordinary,
  title={Ordinary differential equations},
  author={Ince, Edward L},
  year={2012},
  publisher={Courier Corporation}
}

@article{Motohashi:2021zyv,
    author = "Motohashi, Hayato and Noda, Sousuke",
    title = "{Exact solution for wave scattering from black holes: Formulation}",
    eprint = "2103.10802",
    archivePrefix = "arXiv",
    primaryClass = "gr-qc",
    doi = "10.1093/ptep/ptac020",
    journal = "PTEP",
    volume = "2021",
    number = "8",
    pages = "083E03",
    year = "2021"
}

@article{Hatsuda:2021gtn,
    author = "Hatsuda, Yasuyuki and Kimura, Masashi",
    title = "{Spectral Problems for Quasinormal Modes of Black Holes}",
    eprint = "2111.15197",
    archivePrefix = "arXiv",
    primaryClass = "gr-qc",
    reportNumber = "RUP-21-22",
    doi = "10.3390/universe7120476",
    journal = "Universe",
    volume = "7",
    number = "12",
    pages = "476",
    year = "2021"
}

@article{Alday:2009aq,
    author = "Alday, Luis F. and Gaiotto, Davide and Tachikawa, Yuji",
    title = "{Liouville Correlation Functions from Four-dimensional Gauge Theories}",
    eprint = "0906.3219",
    archivePrefix = "arXiv",
    primaryClass = "hep-th",
    doi = "10.1007/s11005-010-0369-5",
    journal = "Lett. Math. Phys.",
    volume = "91",
    pages = "167--197",
    year = "2010"
}

@article{Arnaudo:2024bbd,
    author = "Arnaudo, Paolo and Bonelli, Giulio and Tanzini, Alessandro",
    title = "{One-Loop Corrections to Near-Extremal Kerr Thermodynamics from Semiclassical Virasoro Blocks}",
    eprint = "2412.16057",
    archivePrefix = "arXiv",
    primaryClass = "hep-th",
    doi = "10.1103/cd6l-bl2s",
    journal = "Phys. Rev. Lett.",
    volume = "134",
    number = "25",
    pages = "251401",
    year = "2025"
}

@article{Arnaudo:2025btb,
    author = "Arnaudo, Paolo and Bonelli, Giulio and Tanzini, Alessandro",
    title = "{One loop corrections to the thermodynamics of near-extremal Kerr-(A)dS black holes from Heun equation}",
    eprint = "2506.08959",
    archivePrefix = "arXiv",
    primaryClass = "hep-th",
    month = "6",
    year = "2025"
}

@article{Caceres:2022smh,
    author = "Caceres, Elena and Kundu, Arnab and Patra, Ayan K. and Shashi, Sanjit",
    title = "{Trans-IR flows to black hole singularities}",
    eprint = "2201.06579",
    archivePrefix = "arXiv",
    primaryClass = "hep-th",
    reportNumber = "UTTG-30-2022",
    doi = "10.1103/PhysRevD.106.046005",
    journal = "Phys. Rev. D",
    volume = "106",
    number = "4",
    pages = "046005",
    year = "2022"
}

@article{Caceres:2021fuw,
    author = "Caceres, Elena and Kundu, Arnab and Patra, Ayan K. and Shashi, Sanjit",
    title = "{Page curves and bath deformations}",
    eprint = "2107.00022",
    archivePrefix = "arXiv",
    primaryClass = "hep-th",
    doi = "10.21468/SciPostPhysCore.5.2.033",
    journal = "SciPost Phys. Core",
    volume = "5",
    pages = "033",
    year = "2022"
}

@article{Karch:2025hof,
    author = "Karch, Andreas and Youssef, Merna",
    title = "{Dissipation in Open Holography}",
    eprint = "2509.14312",
    archivePrefix = "arXiv",
    primaryClass = "hep-th",
    reportNumber = "UT-WI-30-2025",
    month = "9",
    year = "2025"
}

@article{Perry:2024vwz,
    author = "Perry, Malcolm and Rodriguez, Maria J.",
    title = "{Love Numbers for Extremal Kerr Black Hole}",
    eprint = "2412.19699",
    archivePrefix = "arXiv",
    primaryClass = "hep-th",
    month = "12",
    year = "2024"
}

@article{Perry:2023wmm,
    author = "Perry, Malcolm and Rodriguez, Maria J.",
    title = "{Dynamical Love Numbers for Kerr Black Holes}",
    eprint = "2310.03660",
    archivePrefix = "arXiv",
    primaryClass = "gr-qc",
    month = "10",
    year = "2023"
}

@article{Chen:2025sbz,
    author = "Chen, Changkai and Jing, Jiliang and Cao, Zhoujian and Wang, Mengjie",
    title = "{Complete quasinormal modes of Type-D black holes}",
    eprint = "2506.14635",
    archivePrefix = "arXiv",
    primaryClass = "gr-qc",
    month = "6",
    year = "2025"
}

@article{Silva:2025khf,
    author = "Silva, Hector O. and Kim, Jung-Wook and Saketh, M. V. S.",
    title = "{Kerr-Newman quasinormal modes and Seiberg-Witten theory}",
    eprint = "2502.17488",
    archivePrefix = "arXiv",
    primaryClass = "gr-qc",
    doi = "10.1103/PhysRevD.111.104021",
    journal = "Phys. Rev. D",
    volume = "111",
    number = "10",
    pages = "104021",
    year = "2025"
}

@article{Bianchi:2021mft,
    author = "Bianchi, Massimo and Consoli, Dario and Grillo, Alfredo and Morales, Jose Francisco",
    title = "{More on the SW-QNM correspondence}",
    eprint = "2109.09804",
    archivePrefix = "arXiv",
    primaryClass = "hep-th",
    doi = "10.1007/JHEP01(2022)024",
    journal = "JHEP",
    volume = "01",
    pages = "024",
    year = "2022"
}

@article{Lei:2023mqx,
    author = "Lei, Yang and Shu, Hongfei and Zhang, Kilar and Zhu, Rui-Dong",
    title = "{Quasinormal modes of C-metric from SCFTs}",
    eprint = "2308.16677",
    archivePrefix = "arXiv",
    primaryClass = "hep-th",
    doi = "10.1007/JHEP02(2024)140",
    journal = "JHEP",
    volume = "02",
    pages = "140",
    year = "2024"
}

@article{Fioravanti:2021dce,
    author = "Fioravanti, Davide and Gregori, Daniele",
    title = "{A new method for exact results on Quasinormal Modes of Black Holes}",
    eprint = "2112.11434",
    archivePrefix = "arXiv",
    primaryClass = "hep-th",
    month = "12",
    year = "2021"
}

@article{Berti:2009kk,
    author = "Berti, Emanuele and Cardoso, Vitor and Starinets, Andrei O.",
    title = "{Quasinormal modes of black holes and black branes}",
    eprint = "0905.2975",
    archivePrefix = "arXiv",
    primaryClass = "gr-qc",
    doi = "10.1088/0264-9381/26/16/163001",
    journal = "Class. Quant. Grav.",
    volume = "26",
    pages = "163001",
    year = "2009"
}

@article{Myers:1986un,
    author = "Myers, Robert C. and Perry, M. J.",
    title = "{Black Holes in Higher Dimensional Space-Times}",
    reportNumber = "PRINT-86-0067 (PRINCETON)",
    doi = "10.1016/0003-4916(86)90186-7",
    journal = "Annals Phys.",
    volume = "172",
    pages = "304",
    year = "1986"
}

@book{conway2012functions,
  title={Functions of one complex variable I},
  author={Conway, John B},
  volume={159},
  year={2012},
  publisher={Springer Science \& Business Media}
}

@article{Suzuki:1999nn,
    author = "Suzuki, Hisao and Takasugi, Eiichi and Umetsu, Hiroshi",
    title = "{Analytic solutions of Teukolsky equation in Kerr-de Sitter and Kerr-Newman-de Sitter geometries}",
    eprint = "gr-qc/9905040",
    archivePrefix = "arXiv",
    reportNumber = "EPHOU-99-007, OU-HET-319",
    doi = "10.1143/PTP.102.253",
    journal = "Prog. Theor. Phys.",
    volume = "102",
    pages = "253--272",
    year = "1999"
}

@article{Suzuki:1998vy,
    author = "Suzuki, Hisao and Takasugi, Eiichi and Umetsu, Hiroshi",
    title = "{Perturbations of Kerr-de Sitter black hole and Heun's equations}",
    eprint = "gr-qc/9805064",
    archivePrefix = "arXiv",
    reportNumber = "EPHOU-98-005, OU-HET-296",
    doi = "10.1143/PTP.100.491",
    journal = "Prog. Theor. Phys.",
    volume = "100",
    pages = "491--505",
    year = "1998"
}

@article{Cvetic:2009jn,
    author = "Cvetic, Mirjam and Larsen, Finn",
    title = "{Greybody Factors and Charges in Kerr/CFT}",
    eprint = "0908.1136",
    archivePrefix = "arXiv",
    primaryClass = "hep-th",
    reportNumber = "UPR-1210-T",
    doi = "10.1088/1126-6708/2009/09/088",
    journal = "JHEP",
    volume = "09",
    pages = "088",
    year = "2009"
}

@article{Cvetic:1997uw,
    author = "Cvetic, Mirjam and Larsen, Finn",
    title = "{General rotating black holes in string theory: Grey body factors and event horizons}",
    eprint = "hep-th/9705192",
    archivePrefix = "arXiv",
    reportNumber = "UPR-0752-T",
    doi = "10.1103/PhysRevD.56.4994",
    journal = "Phys. Rev. D",
    volume = "56",
    pages = "4994--5007",
    year = "1997"
}

@article{Leaver:1986vnb,
    author = "Leaver, E. W.",
    title = "{Solutions to a generalized spheroidal wave equation: Teukolsky{\textquoteright}s equations in general relativity, and the two-center problem in molecular quantum mechanics}",
    doi = "10.1063/1.527130",
    journal = "J. Math. Phys.",
    volume = "27",
    number = "5",
    pages = "1238",
    year = "1986"
}

@article{Hui:2021vcv,
    author = "Hui, Lam and Joyce, Austin and Penco, Riccardo and Santoni, Luca and Solomon, Adam R.",
    title = "{Ladder symmetries of black holes. Implications for love numbers and no-hair theorems}",
    eprint = "2105.01069",
    archivePrefix = "arXiv",
    primaryClass = "hep-th",
    doi = "10.1088/1475-7516/2022/01/032",
    journal = "JCAP",
    volume = "01",
    number = "01",
    pages = "032",
    year = "2022"
}

@book{hitchman2009geometry,
  title={Geometry with an introduction to cosmic topology},
  author={Hitchman, Michael P},
  year={2009},
  publisher={Jones \& Bartlett Learning}
}

@book{lang2013complex,
  title={Complex analysis},
  author={Lang, Serge},
  volume={103},
  year={2013},
  publisher={Springer Science \& Business Media}
}

@book{needham2023visual,
  title={Visual complex analysis},
  author={Needham, Tristan},
  year={2023},
  publisher={Oxford University Press}
}

@article{Gutperle:2025bzr,
    author = "Gutperle, Michael and Yeo, Christina",
    title = "{Janus correlators and Heun's equation}",
    eprint = "2506.01766",
    archivePrefix = "arXiv",
    primaryClass = "hep-th",
    month = "6",
    year = "2025"
}

@article{Jia:2024zes,
    author = "Jia, Hewei Frederic and Rangamani, Mukund",
    title = "{Holographic thermal correlators and quasinormal modes from semiclassical Virasoro blocks}",
    eprint = "2408.05208",
    archivePrefix = "arXiv",
    primaryClass = "hep-th",
    doi = "10.1007/JHEP12(2024)047",
    journal = "JHEP",
    volume = "12",
    pages = "047",
    year = "2024"
}

@article{Consoli:2022eey,
    author = "Consoli, Dario and Fucito, Francesco and Morales, Jose Francisco and Poghossian, Rubik",
    title = "{CFT description of BH{\textquoteright}s and ECO{\textquoteright}s: QNMs, superradiance, echoes and tidal responses}",
    eprint = "2206.09437",
    archivePrefix = "arXiv",
    primaryClass = "hep-th",
    doi = "10.1007/JHEP12(2022)115",
    journal = "JHEP",
    volume = "12",
    pages = "115",
    year = "2022"
}

@article{schafke1980connection,
  title={The connection problem for general linear ordinary differential equations at two regular singular points with applications in the theory of special functions},
  author={Sch{\"a}fke, Reinhard and Schmidt, Dieter},
  journal={SIAM Journal on Mathematical Analysis},
  volume={11},
  number={5},
  pages={848--862},
  year={1980},
  publisher={SIAM}
}

@article{darboux1878memoire,
  title={M{\'e}moire sur l'approximation des fonctions de tr{\`e}s-grands nombres, et sur une classe {\'e}tendue de d{\'e}veloppements en s{\'e}rie},
  author={Darboux, Gaston},
  journal={Journal de Math{\'e}matiques pures et appliqu{\'e}es},
  volume={4},
  pages={5--56},
  year={1878}
}

@article{Bonelli:2022ten,
    author = "Bonelli, Giulio and Iossa, Cristoforo and Panea Lichtig, Daniel and Tanzini, Alessandro",
    title = "{Irregular Liouville Correlators and Connection Formulae for Heun Functions}",
    eprint = "2201.04491",
    archivePrefix = "arXiv",
    primaryClass = "hep-th",
    doi = "10.1007/s00220-022-04497-5",
    journal = "Commun. Math. Phys.",
    volume = "397",
    number = "2",
    pages = "635--727",
    year = "2023"
}

@article{Bonelli:2021uvf,
    author = "Bonelli, Giulio and Iossa, Cristoforo and Lichtig, Daniel Panea and Tanzini, Alessandro",
    title = "{Exact solution of Kerr black hole perturbations via CFT2 and instanton counting: Greybody factor, quasinormal modes, and Love numbers}",
    eprint = "2105.04483",
    archivePrefix = "arXiv",
    primaryClass = "hep-th",
    doi = "10.1103/PhysRevD.105.044047",
    journal = "Phys. Rev. D",
    volume = "105",
    number = "4",
    pages = "044047",
    year = "2022"
}

@article{Derezi_ski_2013,
   title={Hypergeometric Type Functions and Their Symmetries},
   volume={15},
   ISSN={1424-0661},
   url={http://dx.doi.org/10.1007/s00023-013-0282-4},
   DOI={10.1007/s00023-013-0282-4},
   number={8},
   journal={Annales Henri Poincaré},
   publisher={Springer Science and Business Media LLC},
   author={Dereziński, Jan},
   year={2013},
   month=sep, pages={1569–1653} }

@article{maier2007192,
  title={The 192 solutions of the Heun equation},
  author={Maier, Robert},
  journal={Mathematics of Computation},
  volume={76},
  number={258},
  pages={811--843},
  year={2007}
}

@article{Gray:2024qys,
    author = "Gray, Finnian and Keeler, Cynthia and Kubiznak, David and Martin, Victoria",
    title = "{Love symmetry in higher-dimensional rotating black hole spacetimes}",
    eprint = "2409.05964",
    archivePrefix = "arXiv",
    primaryClass = "gr-qc",
    doi = "10.1007/JHEP03(2025)036",
    journal = "JHEP",
    volume = "03",
    pages = "036",
    year = "2025"
}

@article{Charalambous:2022rre,
    author = "Charalambous, Panagiotis and Dubovsky, Sergei and Ivanov, Mikhail M.",
    title = "{Love symmetry}",
    eprint = "2209.02091",
    archivePrefix = "arXiv",
    primaryClass = "hep-th",
    doi = "10.1007/JHEP10(2022)175",
    journal = "JHEP",
    volume = "10",
    pages = "175",
    year = "2022"
}

@article{Charalambous:2021kcz,
    author = "Charalambous, Panagiotis and Dubovsky, Sergei and Ivanov, Mikhail M.",
    title = "{Hidden Symmetry of Vanishing Love Numbers}",
    eprint = "2103.01234",
    archivePrefix = "arXiv",
    primaryClass = "hep-th",
    reportNumber = "INR-TH-2021-003",
    doi = "10.1103/PhysRevLett.127.101101",
    journal = "Phys. Rev. Lett.",
    volume = "127",
    number = "10",
    pages = "101101",
    year = "2021"
}

@article{Charalambous:2023jgq,
    author = "Charalambous, Panagiotis and Ivanov, Mikhail M.",
    title = "{Scalar Love numbers and Love symmetries of 5-dimensional Myers-Perry black holes}",
    eprint = "2303.16036",
    archivePrefix = "arXiv",
    primaryClass = "hep-th",
    doi = "10.1007/JHEP07(2023)222",
    journal = "JHEP",
    volume = "07",
    pages = "222",
    year = "2023"
}

@article{Charalambous:2021mea,
    author = "Charalambous, Panagiotis and Dubovsky, Sergei and Ivanov, Mikhail M.",
    title = "{On the Vanishing of Love Numbers for Kerr Black Holes}",
    eprint = "2102.08917",
    archivePrefix = "arXiv",
    primaryClass = "hep-th",
    reportNumber = "INR-TH-2021-001",
    doi = "10.1007/JHEP05(2021)038",
    journal = "JHEP",
    volume = "05",
    pages = "038",
    year = "2021"
}

@article{Kol:2011vg,
    author = "Kol, Barak and Smolkin, Michael",
    title = "{Black hole stereotyping: Induced gravito-static polarization}",
    eprint = "1110.3764",
    archivePrefix = "arXiv",
    primaryClass = "hep-th",
    doi = "10.1007/JHEP02(2012)010",
    journal = "JHEP",
    volume = "02",
    pages = "010",
    year = "2012"
}

@article{Hui:2020xxx,
    author = "Hui, Lam and Joyce, Austin and Penco, Riccardo and Santoni, Luca and Solomon, Adam R.",
    title = "{Static response and Love numbers of Schwarzschild black holes}",
    eprint = "2010.00593",
    archivePrefix = "arXiv",
    primaryClass = "hep-th",
    doi = "10.1088/1475-7516/2021/04/052",
    journal = "JCAP",
    volume = "04",
    pages = "052",
    year = "2021"
}

@article{Glazer:2024eyi,
    author = "Glazer, Daniel and Joyce, Austin and Rodriguez, Maria J. and Santoni, Luca and Solomon, Adam R. and Temoche, Luis Fernando",
    title = "{Higher-Dimensional Black Holes and Effective Field Theory}",
    eprint = "2412.21090",
    archivePrefix = "arXiv",
    primaryClass = "hep-th",
    month = "12",
    year = "2024"
}

@article{Noda:2022zgk,
    author = "Noda, Sousuke and Motohashi, Hayato",
    title = "{Spectroscopy of Kerr-AdS5 spacetime with the Heun function: Quasinormal modes, greybody factor, and evaporation}",
    eprint = "2206.07721",
    archivePrefix = "arXiv",
    primaryClass = "gr-qc",
    doi = "10.1103/PhysRevD.106.064025",
    journal = "Phys. Rev. D",
    volume = "106",
    number = "6",
    pages = "064025",
    year = "2022"
}

@article{Sasaki:2003xr,
    author = "Sasaki, Misao and Tagoshi, Hideyuki",
    title = "{Analytic black hole perturbation approach to gravitational radiation}",
    eprint = "gr-qc/0306120",
    archivePrefix = "arXiv",
    doi = "10.12942/lrr-2003-6",
    journal = "Living Rev. Rel.",
    volume = "6",
    pages = "6",
    year = "2003"
}

@book{ronveaux1995heun,
  title={Heun's differential equations},
  author={Ronveaux, Andr{\'e} and Arscott, Felix Medland},
  year={1995},
  publisher={Clarendon Press}
}

@article{Lisovyy:2022flm,
    author = "Lisovyy, O. and Naidiuk, A.",
    title = "{Perturbative connection formulas for Heun equations}",
    eprint = "2208.01604",
    archivePrefix = "arXiv",
    primaryClass = "math-ph",
    doi = "10.1088/1751-8121/ac9ba7",
    journal = "J. Phys. A",
    volume = "55",
    number = "43",
    pages = "434005",
    year = "2022"
}

@article{Aminov:2023jve,
    author = "Aminov, Gleb and Arnaudo, Paolo and Bonelli, Giulio and Grassi, Alba and Tanzini, Alessandro",
    title = "{Black hole perturbation theory and multiple polylogarithms}",
    eprint = "2307.10141",
    archivePrefix = "arXiv",
    primaryClass = "hep-th",
    doi = "10.1007/JHEP11(2023)059",
    journal = "JHEP",
    volume = "11",
    pages = "059",
    year = "2023"
}

@article{Jorge:2014kra,
    author = "Jorge, Rog\'erio and de Oliveira, Ednilton S. and Rocha, Jorge V.",
    title = "{Greybody factors for rotating black holes in higher dimensions}",
    eprint = "1410.4590",
    archivePrefix = "arXiv",
    primaryClass = "gr-qc",
    doi = "10.1088/0264-9381/32/6/065008",
    journal = "Class. Quant. Grav.",
    volume = "32",
    number = "6",
    pages = "065008",
    year = "2015"
}

@article{MyersPerry,
    author  = {Myers, Robert C. and Perry, Malcolm J.},
    title   = {Black Holes in Higher Dimensional Space-Times},
    journal = {Annals of Physics},
    volume  = {172},
    number  = {2},
    pages   = {304-347},
    year    = {1986},
    doi     = {10.1016/0003-4916(86)90186-7},
    eprint  = {arXiv:1001.4527},
    archivePrefix = {arXiv},
    primaryClass = {gr-qc}
}

@article{Dias:2010eu,
    author = "Dias, Oscar J. C. and Figueras, Pau and Monteiro, Ricardo and Reall, Harvey S. and Santos, Jorge E.",
    title = "{An instability of higher-dimensional rotating black holes}",
    eprint = "1001.4527",
    archivePrefix = "arXiv",
    primaryClass = "hep-th",
    doi = "10.1007/JHEP05(2010)076",
    journal = "JHEP",
    volume = "05",
    pages = "076",
    year = "2010"
}

@article{Guica:2008mu,
    author = "Guica, Monica and Hartman, Thomas and Song, Wei and Strominger, Andrew",
    title = "{The Kerr/CFT Correspondence}",
    eprint = "0809.4266",
    archivePrefix = "arXiv",
    primaryClass = "hep-th",
    doi = "10.1103/PhysRevD.80.124008",
    journal = "Phys. Rev. D",
    volume = "80",
    pages = "124008",
    year = "2009"
}

@article{LeTiec:2020bos,
    author = "Le Tiec, Alexandre and Casals, Marc and Franzin, Edgardo",
    title = "{Tidal Love Numbers of Kerr Black Holes}",
    eprint = "2010.15795",
    archivePrefix = "arXiv",
    primaryClass = "gr-qc",
    doi = "10.1103/PhysRevD.103.084021",
    journal = "Phys. Rev. D",
    volume = "103",
    number = "8",
    pages = "084021",
    year = "2021"
}

@article{Hui:2022vbh,
    author = "Hui, Lam and Joyce, Austin and Penco, Riccardo and Santoni, Luca and Solomon, Adam R.",
    title = "{Near-zone symmetries of Kerr black holes}",
    eprint = "2203.08832",
    archivePrefix = "arXiv",
    primaryClass = "hep-th",
    doi = "10.1007/JHEP09(2022)049",
    journal = "JHEP",
    volume = "09",
    pages = "049",
    year = "2022"
}

@book{Bateman:100233,
      author        = "Bateman, Harry and Erdélyi, Arthur",
      title         = "{Higher transcendental functions}",
      publisher     = "McGraw-Hill",
      address       = "New York, NY",
      series        = "Calif. Inst. Technol. Bateman Manuscr. Project",
      year          = "1955",
      url           = "https://cds.cern.ch/record/100233",
}

@book{slavjanov2000special,
  title={Special Functions: A Unified Theory Based on Singularities},
  author={Slavjanov, S.J. and Wolfgang, L.},
  isbn={9780198505730},
  lccn={00038564},
  series={Oxford Science Publications},
  year={2000},
  publisher={Oxford University Press}
}

@article{Rodriguez:2023xjd,
    author = "Rodriguez, Maria J. and Santoni, Luca and Solomon, Adam R. and Temoche, Luis Fernando",
    title = "{Love numbers for rotating black holes in higher dimensions}",
    eprint = "2304.03743",
    archivePrefix = "arXiv",
    primaryClass = "hep-th",
    doi = "10.1103/PhysRevD.108.084011",
    journal = "Phys. Rev. D",
    volume = "108",
    number = "8",
    pages = "084011",
    year = "2023"
}
\bibliographystyle{utphys}
\end{document}